\documentclass{aa}  

\usepackage{graphicx}
\usepackage{txfonts}
\usepackage{xspace}
\usepackage[T1]{fontenc}

\usepackage{xcolor}

\newcommand{\nnhp}[0]{N$_2$H$^+$\xspace}
\newcommand{\hcop}[0]{HCO$^+$\xspace}
\newcommand{\tco}[0]{$^{13}$CO\xspace}
\newcommand{\ceto}[0]{C$^{18}$O\xspace}
\newcommand{\cch}[0]{C$_{2}$H\xspace}
\newcommand{\CO}[0]{$^{12}$CO\xspace}

\usepackage{natbib}
\bibpunct{(}{)}{;}{a}{}{,}

\usepackage{hyperref}
\begin{document}

    \title{Surveying the Whirlpool at Arcseconds with NOEMA (SWAN)}
    \subtitle{II: Survey design and observations}

\author{Sophia K. Stuber\inst{\ref{MPIA},\ref{UniHD}}
          \and Jerome Pety\inst{\ref{IRAM},\ref{LERMA}}
          \and Antonio Usero\inst{\ref{OAN}}
          \and Eva Schinnerer\inst{\ref{MPIA}}
          \and Frank Bigiel\inst{\ref{AlfA}}
          \and Mar\'ia~J.~Jim\'enez-Donaire\inst{\ref{OAN},\ref{Yebes}}
          \and Jakob den Brok\inst{\ref{CfA}}
          \and Adam K. Leroy \inst{\ref{OSUAstro}}
          \and Ina Gali\'{c}\inst{\ref{AlfA}}
          \and Annie~Hughes\inst{\ref{IRAP}}
          \and Mallory Thorp\inst{\ref{AlfA}}
          \and Ashley.~T.~Barnes\inst{\ref{ESO}}
          \and Ivana Be\v{s}li\'{c}\inst{\ref{LERMA}}
          \and Cosima Eibensteiner \inst{\ref{NRAO}}\thanks{Jansky Fellow of the National Radio Astronomy Observatory}
          \and Damian R. Gleis\inst{\ref{MPIA}}
          \and Ralf~S.~Klessen\inst{\ref{ITA},\ref{IWR},\ref{CfA},\ref{Rad}}
          \and Daizhong Liu\inst{\ref{PMO}} 
          \and Hsi-An Pan \inst{\ref{TKU}}
          \and Toshiki Saito\inst{\ref{Shizuoka}}
          \and Sumit K. Sarbadhicary \inst{\ref{OSUAstro}}
          \and Thomas~G.~Williams\inst{\ref{Ox}}
          }

\institute{
    Max-Planck-Institut für Astronomie, Königstuhl 17, 69117 Heidelberg, Germany\label{MPIA}
    \and 
    Universit\"{a}t Heidelberg, Astronomy and Physics department, 69120 Heidelberg, Germany \label{UniHD}
    \and 
    IRAM, 300 rue de la Piscine, 38400 Saint Martin d'H\`eres, France\label{IRAM}
    \and 
    Sorbonne Universit\'e, Observatoire de Paris, Universit\'e PSL, \' Ecole normale sup\`erieure, CNRS, LERMA, F-75005, Paris, France\label{LERMA}
    \and    
    Observatorio Astron\'omico Nacional (IGN), C/ Alfonso XII, 3, E-28014 Madrid, Spain\label{OAN}
    \and 
    Argelander-Institut für Astronomie, Universität Bonn, Auf dem Hügel 71, 53121 Bonn, Germany\label{AlfA}
    \and
    Centro de Desarrollos Tecnol\'ogicos, Observatorio de Yebes (IGN), 19141 Yebes, Guadalajara, Spain\label{Yebes}
    \and 
    Center for Astrophysics $\mid$ Harvard \& Smithsonian, 60 Garden St., 02138 Cambridge, MA, USA\label{CfA}
    \and 
    Department of Astronomy, The Ohio State University, Columbus, Ohio 43210, USA \label{OSUAstro}
    \and
    IRAP, OMP, Université de Toulouse, 9 avenue du Colonel Roche, Toulouse 31028 cedex 4, France \label{IRAP}
    \and 
    European Southern Observatory, Karl-Schwarzschild 2, 85748 Garching bei Muenchen, Germany\label{ESO}
    \and 
    National Radio Astronomy Observatory, 520 Edgemont Road, Charlottesville, VA 22903, USA\label{NRAO}
    \and 
    Universit\"{a}t Heidelberg, Zentrum f\"{u}r Astronomie, Institut f\"{u}r Theoretische Astrophysik, Albert-Ueberle-Str.\ 2, 69120 Heidelberg, Germany\label{ITA}
    \and
    Universit\"{a}t Heidelberg, Interdisziplin\"{a}res Zentrum f\"{u}r Wissenschaftliches Rechnen, Im Neuenheimer Feld 225, 69120 Heidelberg, Germany\label{IWR}
    \and
    Radcliffe Institute for Advanced Study, Harvard University,  10 Garden St, 02138 Cambridge, MA, USA \label{Rad}
    \and
    Purple Mountain Observatory, Chinese Academy of Sciences, 10 Yuanhua Road, Nanjing 210023, China\label{PMO}
    \and
    Department of Physics, Tamkang University, No.151, Yingzhuan Road, Tamsui District, New Taipei City 251301, Taiwan \label{TKU}
    \and
    Faculty of Global Interdisciplinary Science and Innovation, Shizuoka University, 836 Ohya, Suruga-ku, Shizuoka 422-8529, Japan\label{Shizuoka}
    \and
    Sub-department of Astrophysics, Department of Physics, University of Oxford, Keble Road, Oxford OX1 3RH, UK\label{Ox}
}

   \date{March 2025}

  \abstract{
    We present Surveying the Whirlpool at Arcseconds with NOEMA (SWAN), a high-resolution, high-sensitivity survey to map molecular lines in the 3mm band in M51 (the Whirlpool galaxy).
    SWAN has obtained the largest high-sensitivity map ($\sim 5\times 7\,$kpc$^2$) of \nnhp emission at $\sim$cloud-scale resolution ($3\arcsec\sim 125\,$pc) in an external galaxy to date.
    Here we describe the observations and data reduction of $\sim214\,$hours of interferometric data from the Northern Extended Millimetre Array (NOEMA)  and $\sim55$\,hours of tailored new observations with the 30m telescope of the Institut de radioastronomie millimétrique (IRAM) as well as the combination of the NOEMA, new 30m and $\sim14$\,hours of archival IRAM-30m observations. 
    We detect widespread emission from nine molecular transition lines. 
    The J=1-0 transitions of the CO isotopologues \tco and \ceto are detected at high significance across the full observed field-of-view (FoV). 
    HCN(1-0), HNC(1-0), \hcop{}(1-0), and \nnhp{}(1-0) are detected in the center, molecular ring and spiral arms of the galaxy, while the shock tracer HNCO(4-3) and (5-4) and PDR tracer \cch{}(1-0) are detected in the central $\sim$1 kpc and molecular ring only. 
    For most of the lines that we detect, average line ratios with respect to \CO are increased by up to a factor of $\sim3$ in the central 1\,kpc, where an AGN and its low-inclination outflow are present, compared to the disk. 
    Line ratios between CO isotopologues show less variation across the SWAN FoV.
    Across the full SWAN FoV, \tco, \ceto, HCN, HNC, \hcop and \nnhp are $8\pm^2_2$, $29\pm^{7}_{6}$, $17\pm^3_5$, $37\pm^5_{10}$, $26\pm^5_3$ and $63\pm^{38}_{10}$ times fainter than \CO, respectively, 
    in pixels where each line is significantly detected.
    Although we observe variations in line ratios between larger-scale environments like the center and disk of M51, the scatter within each environment also indicates the influence of smaller-scale processes. 
    The ability to measure these effects is only possible thanks to the high resolution and high sensitivity of the SWAN dataset across multiple environments. 
    This provides the sharpest view of these molecular transitions over the largest physical area ever captured in an external galaxy.
    }

   \maketitle

\section{Introduction}
\label{sec:Introduction}

Star formation is a fundamental process for mass growth of galaxies and thus their evolution. While the rate of star formation is closely tied to the amount of molecular gas present on scales ranging from individual cores inside molecular clouds to entire galaxies  \citep[e.g.,][]{bigiel_star_2008,kennicutt_star_2012}, (massive) star formation occurs only in the densest regions of molecular clouds \citep{gao_star_2004,wu_connecting_2005, lada_star_2012,evans_2014}. 
These dense regions within molecular clouds are generally too small to be resolved at extragalactic distances. 
Therefore, to study molecular gas at different densities, especially in external galaxies, astronomers combine emission lines that span a wide range of critical or effective excitation densities \citep[e.g.,][]{shirley_critical_2015}.

The high-dipole moment molecules HCN, its isomer HNC, and \hcop all have high critical densities and relatively bright rotational transitions in the $\lambda = 3$~mm window. As a result, these emission lines have often been considered as ``dense gas tracers'' and observed in extragalactic targets for decades \citep{helfer_dense_1993,aalto_variation_1997,aalto_cn_2002,gao_star_2004, brouillet_hcn_2005,aalto_detection_2012, buchbender_dense_2013, usero_variations_2015,neumann_almond_2023, bigiel_empire_2016, jimenez-donaire_empire_2019, krieger_molecular_2020,beslic_dense_2021,eibensteiner_23_2022, rybak_prussic_2022,imanishi_dense_2023}.
Many of these surveys reveal systematic variations in line ratios like HCN/CO as a function of environment \citep[see references above and review by][]{schinnerer_molecular_2024}, suggesting that the host galaxy sets the initial conditions for star formation.  
Even further, the efficiency with which stars are formed out of dense gas, seems to vary drastically within larger-scale structures, such as larger-scale bars \citep{beslic_dense_2021,eibensteiner_23_2022,neumann_ngc4321_2024} or even our own Milky Way’s Central Molecular Zone \citep[CMZ,][]{longmore_variations_2013}.

Unfortunately, these emission lines are also typically $\gtrsim20$ times fainter than \CO \citep{usero_variations_2015,jimenez-donaire_empire_2019, jimenez-donaire_constant_2023,schinnerer_molecular_2024}. 
As a result, recent observations of those dense gas tracers attempting to link galaxy environment and molecular gas conditions focus either on mapping larger regions of galaxy disks at low-resolution \citep[$\sim\,500-1000$pc;][]{bigiel_empire_2016, jimenez-donaire_empire_2019, gallagher_dense_2018, heyer_dense_2022,neumann_almond_2023} or focus at higher-resolution (i.e., $\sim100\,$pc) observations of individual regions within galaxies \citep[e.g., galaxy centers, spiral arms;][]{chen_dense_2017, beslic_dense_2021,beslic_properties_2024, neumann_ngc4321_2024}. 
While the former is insufficient to isolate individual star-forming regions of molecular clouds, the latter does not capture changes across different environments. 

Clearly, a major next step in dense gas tracer studies is to observe these tracers at both much higher physical resolution and across larger areas in galaxy disks. 
This paper presents the IRAM Large Program ``Surveying the Whirlpool at Arcseconds with NOEMA'' (SWAN), which aims to take this natural next step. 
SWAN used the NOrthern Extended Millimetre Array (NOEMA) and IRAM 30m single dish telescope to map the emission from dense gas tracers (HCN(1-0), HNC(1-0), \hcop(1-0), \nnhp(1-0)) in the 3mm band across a  large $5\times7\,$kpc$^2$ portion of the prototypical grand-design spiral M51, one of the closest \citep[$D\approx8.5$~Mpc;][]{mcquinn_distance_2016} northern, face-on, star-forming galaxies. 
Additionally, we observe CO isotopologues \ceto, \tco{}(1-0), shock-tracing emission lines (HNCO(4-3), HNCO(5-4)), and tracers of photo-dissociation regions (PDR, C$_2$H(1-0)). 
We achieve $125$~pc resolution, sufficient to resolve the population of giant molecular clouds (GMCs) and approaching the size scale of individual massive GMCs or star-forming complexes.

Of particular interest is the first high resolution, high sensitivity wide-field extragalactic map of \nnhp from SWAN \citep{stuber_surveying_2023}. 
Galactic studies, which detect a much larger suite of molecular emission lines due to the proximity of their targets, prefer the use of this molecular ion, \nnhp, over HCN, HNC and \hcop to identify regions of dense ($n\sim10^5$\,cm$^{-3}$) molecular gas.
Based on Galactic studies, there is active ongoing discussion in the literature about how the observed intensity of HCN, HNC and \hcop depends on the gas density distribution, chemical abundances, and other factors \citep[][]{shirley_critical_2015, gallagher_spectroscopic_2018, pety_anatomy_2017,kauffmann_molecular_2017,leroy_millimeter-wave_2017,heyer_dense_2022,santa-maria_hcn_2023,neumann_almond_2023,tafalla_characterizing_2023}.
\nnhp, in contrast, not only has a high critical density, but chemical reactions in the molecular phase of the interstellar medium (ISM) ensure that \nnhp is a selective tracer of dense gas. 
One of the main destruction mechanisms of \nnhp is the reaction with \CO to form \hcop and other molecules. 
\nnhp therefore exclusively survives in the densest, coldest parts of molecular clouds, where \CO is frozen out onto dust grains \citep[column densities above 10$^{22}$ cm$^{-2}$, see][]{pety_anatomy_2017,kauffmann_molecular_2017, tafalla_characterizing_2021}. 
Since \nnhp emission is more than $\sim70$ times fainter than \CO emission \citep[e.g.,][]{jimenez-donaire_constant_2023,stuber_surveying_2023}, it has been challenging to observe in extragalactic targets. Previous observations of \nnhp in targets beyond the Local Group include a handful of single-target single-dish studies with kpc-scale resolution \citep[e.g.,][]{den_brok_co_2022, jimenez-donaire_constant_2023} and a few  dedicated higher resolution studies of bright galaxy centers \citep[e.g.,][]{meierturner2005, martin_alchemi_2021, eibensteiner_23_2022}. 
In light of the importance of mapping this tracer across a galaxy disk, preliminary maps of \nnhp and HCN from the SWAN survey have already been presented in \citet[][hereafter: S23]{stuber_surveying_2023}.

SWAN builds on the high quality view of the molecular ISM from CO~(1-0) mapping by PAWS \citep[][]{schinnerer_pdbi_2013,meidt_gas_2013,colombo_pdbi_2014} as well as overlapping coverage by VLA radio continuum mapping (including free-free emission), JWST and HST recombination line and infrared mapping \citep[e.g.,][]{robert_c_kennicutt_star_2007,dumas_local_2011, querejeta_dense_2019, kessler_pa_2020}. 
These means that the bulk molecular ISM and recent star formation are uniquely well-constrained in this galaxy. The ISM and dynamical environment across the galaxy are also well-understood from extensive previous multiwavelength analysis \citep[e.g.,][]{walter_things_2008,meidt_gas_2013,colombo_pdbi_2014-1,den_brok_co_2022}. 
M51 was targeted by several previous lower resolution dense gas tracer mapping studies that detected strong environmental variations in, e.g., the HCN/CO ratio across the galaxy \citep[at 500-1000pc resolution][]{usero_variations_2015,bigiel_empire_2016,heyer_dense_2022}. 
At $\sim\,100$pc resolution, individual regions within M51 were targeted \citep[i.e.,][]{chen_dense_2017,querejeta_dense_2019}.
This makes M51 an ideal target to investigate the physical origin of the environmental variations of HCN/CO, SFR/HCN, and similar quantities observed at much larger spatial scale.

This paper presents the full suite of 3\,mm lines observed as part of the SWAN IRAM Large Program, and is structured as follows.  In Section~\ref{sec:Observations}, we describe the SWAN observations.  The data reduction, imaging and data quality assessment are presented in Section~\ref{sec:Datareduction}. 
In Section~\ref{sec:Results}, we compare the spatial distribution of all the nine 3\,mm lines detected in the SWAN data set, and compare their relative intensity to \CO and previous literature observations. An interpretation of these findings is provided in Section~\ref{sec:Discussion}. We summarize our main conclusions in Section~\ref{sec:Summary}.
The SWAN data is publicly available on the IRAM website\footnote{\url{https://oms.iram.fr/dms}}. 

The richness of the SWAN data allows for more detailed studies of, e.g., the variations of \CO isotopologues \citep{Galic_submitted} also in combination with SMA observations in \citet{den_Brok_2024SWAN_SMA}, a direct comparison of the dense gas tracing lines across different environments \citep[for a first comparison of HCN and \nnhp, see][with an extensive comparison including HNC and \hcop underway; Stuber et al. in prep.]{stuber_surveying_2023} and to cloud properties (e.g., gas mass surface density, line width, virial parameter; Bigiel et al., in prep.). Further dedicated studies of the central molecular outflow are under way (Thorp et al. in prep., Usero et al., in prep.).

\section{Observations}
\label{sec:Observations}

SWAN utilizes observations from the IRAM Large Program LP003 (PIs: E. Schinnerer \& F. Bigiel), which mapped emission lines in the $\sim3\,$mm band across the central 5$\times$7\,kpc$^2$ of the nearby galaxy M51 with both the NOEMA and the IRAM 30m single dish telescope.

\subsection{NOEMA observations }
\label{sec:NOEMAobs}

\begin{figure}
    \centering
    \includegraphics[width = 0.24\textwidth]{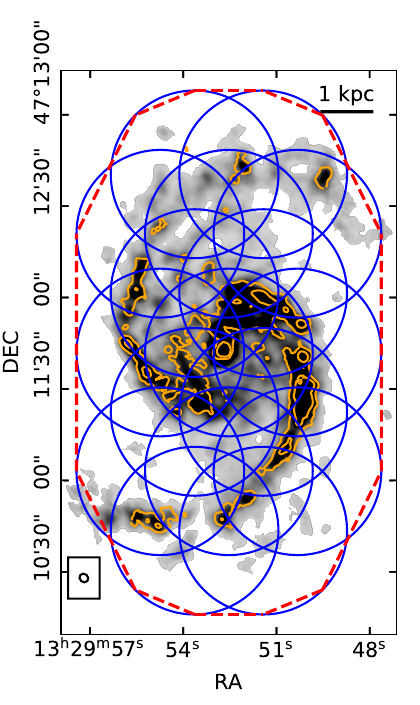}
     \includegraphics[width = 0.24\textwidth]{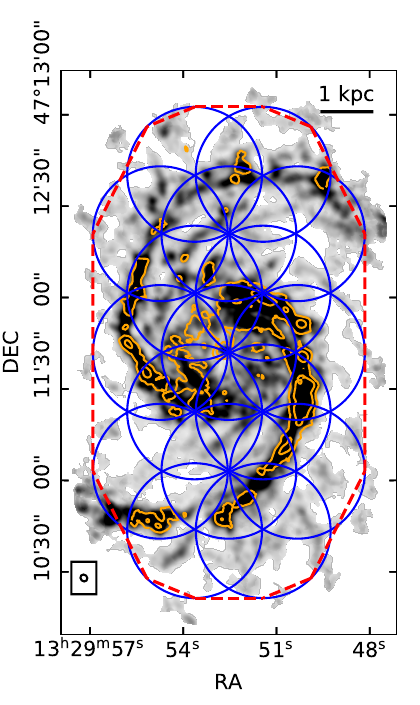}
    \caption{Primary beam of individual mosaic pointings (blue circles) of the NOEMA observations for SWAN overlaid on a map of the integrated intensity of HCN(1-0) emission (left) and \tco{}(1-0) emission (right).
    Contours represent integrated \nnhp emission at 0.5 and 2 K km\,s$^{-1}$. The pointings shown have a radius of $\sim28\arcsec$ (HCN) and $\sim22.5\arcsec$ (\tco, our highest frequency detected line). 
    We interpolate the outer edges of the individual mosaic pointings to define the area inside which we measure the integrated flux of each observed emission line (red line). Because the primary beam changes with frequency, so does the area used to calculate the line flux.}
    \label{fig:mosaicpointings}
\end{figure}

Observations with NOEMA were taken between 2020 January and 2021 December. 
with the antenna configuration split between the C (59\%; 126h) and D (41\%; 88h) configurations.
In total, we used 17 pointings to map the central $\rm \sim5\times7\,kpc^2$ of M51's disk (Figure~\ref{fig:mosaicpointings}). The mosaic uses hexagonal spacing.

For the SWAN observations, the receiver was tuned to a sky frequency of 88.49177 GHz, which corresponds to the frequency of HCN(1-0) redshifted by the systemic velocity of M51 ($v_\mathrm{sys} \sim 471.7\,$km\,s$^{-1}$). We covered the $\sim15.5\,$GHz ($\times$2 polarisations) instantaneous bandwidth at the default Polyfix 2\,MHz resolution. 
The setup was selected to cover the main target lines \tco, \ceto(1-0) in the upper side band, as well as HCN, \hcop HNC and \nnhp (1-0) in the lower side band  (Table~\ref{tab:Dataquality}). 
As the configuration of Polyfix permits up to 16 high-resolution (62.5\,kHz) windows per 4\,GHz, we placed between three and five such windows around the rest frequency of each of our target lines.

We evaluated the quality of the SWAN observations based on the automatic calibration reports obtained with the standard \texttt{GILDAS/CLIC} calibration pipeline.  The automatic quality assessment tool was used to filter out poor data (see \texttt{GILDAS/CLIC} manual\footnote{\url{https://www.iram.fr/IRAMFR/GILDAS/doc/html/clic-html/clic.html}}). Out of a total of $\sim246$h of observations, $\sim214$h were taken under average to excellent observing conditions (i.e., low water vapour, minimum cloud coverage, minimum antenna tracking errors without systematic variations, good phase and amplitude stability). 
The average water vapor during these 214h was $\sim 4\,$mm.

\begin{figure}[h]
    \centering
    \includegraphics[width = 0.47\textwidth]{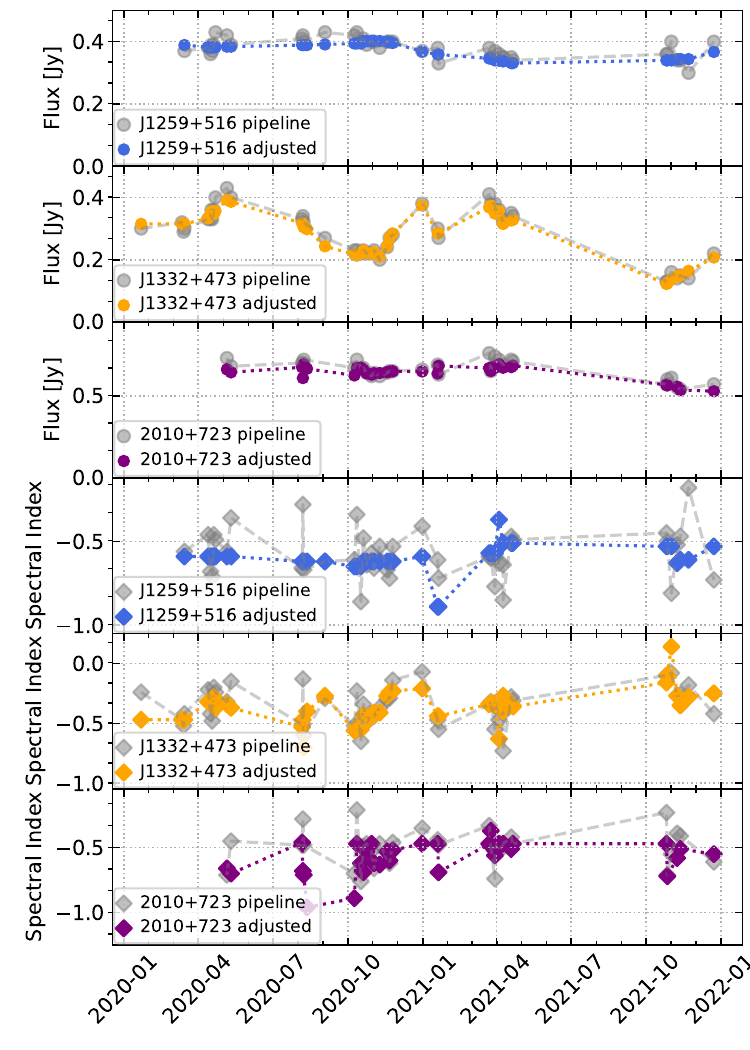}
    \caption{Observed flux (top three rows) and spectral index (bottom three rows) over time for the three used phase and amplitude calibrators of the SWAN NOEMA observations. In addition to the adopted solution  (colored points), we show the flux solution determined by the pipeline (grey circles). The temporal variation in the flux of these quasars is relatively smooth.
    Manual adjustments result in smaller time variations of the spectral index.}
    \label{fig:FluxcalibrationNOEMA}
\end{figure}

We used the quasars J1259+516 and J1332+473 as our main phase and amplitude calibrators, substituted by 1418+546 if either calibration target was unavailable. 
Observations of the calibrators were executed every $\sim17\,$minutes. 
Absolute flux calibration was performed using IRAM models for MWC349 and LkHa101, providing about 50 independent measurements of the flux of J1259+516 and J1332+473 and quasar 2010+723 over a period of about one year. This allowed us to confirm that variations of the flux of these quasars happen mostly over longer time periods and were relatively smooth.
We show the flux of J1259+516, J1332+473 and 2010+723 over time in Figure~\ref{fig:FluxcalibrationNOEMA}, inferred by the \texttt{GILDAS} pipeline solution, and after manual adjustments. While the pipeline is deriving the flux gain for each observation separately by ensuring that the flux of a primary flux calibrator is equal to the modeled value, the manual solution uses  the temporal stability of the most stable quasar from one day to the next. 
That means we take advantage of the fact that observations were executed over consecutive days. For sets of observations taken on consecutive days, we identify as new flux reference the quasar whose flux is most stable over this time range. 
We use its derived flux value from the day with best observing conditions as reference to solve again for the flux gains of the other observations in each set. 
Significant changes in the derived flux gains only occurred when the observations of the primary flux calibrator were not taken under good conditions, e.g. the elevation of the primary flux calibrator was low. Since the full mosaic is observed between two phase calibrator observations, the flux calibration has the same effect for all its 17 pointings. The same is true for all lines, as they are observed simultaneously and a global spectral factor is applied per sideband. 
In short, while the flux calibration impacts the absolute fluxes of the SWAN data, it does not impact the relative strengths between the different lines observed.
Based on Figure~\ref{fig:FluxcalibrationNOEMA}, we estimate that the absolute flux uncertainty for the NOEMA observations is $\sim10\%$.

\begin{table*}
\begin{small}
\caption{Overview of NOEMA and 30m observations}\label{tab:ObservationalSettings}
\centering
\begin{tabular}{l|ll|llllll}
\hline\hline
\noalign{\smallskip}
    Survey &  Telescope & Lines & $\Delta t_\mathrm{obs}$ (in SWAN FoV) &  native resolution  \\
\noalign{\smallskip}
\hline
\noalign{\smallskip}
    SWAN& NOEMA & \cch{}(1-0), HNCO(4-3), HCN(1-0), \hcop{}(1-0), &  $214$h & $2.4\arcsec{}- 3.0\arcsec$\\
     &  &HNC(1-0), \nnhp{}(1-0), \ceto{}(1-0), HNCO(5-4), & & \\
     &  & \tco{}(1-0)  & & \\
\noalign{\smallskip}
\hline
\noalign{\smallskip}
    SWAN-new & IRAM-30m  & same as SWAN-NOEMA & $55$h & $ 23.6\arcsec{} - 29.8\arcsec$  \\
    CLAWS & IRAM-30m & \ceto{}(1-0), \tco{}(1-0), \nnhp{}(1-0) & $\sim14$h & $27-32\arcsec$\\
    EMPIRE & IRAM-30m & HCN(1-0), HNC(1-0), \hcop{}(1-0) & $\sim14$h & $26-34\arcsec$\\
    \noalign{\smallskip}
\end{tabular}
\tablefoot{Observational data used in the SWAN survey, including archival data from EMPIRE \citep{jimenez-donaire_empire_2019} and CLAWS \citep{den_brok_co_2022}. 
The observing time $t_\mathrm{obs}$ refers to the estimated observing time inside the FoV of SWAN (see Figure~\ref{fig:mosaicpointings}). All 30m datasets (SWAN-new, EMPIRE, CLAWS) are combined to the ``SWAN-30m'' data set before merging with the NOEMA data via joint deconvolution. Native resolution refers to the resolution after imaging. } 
\end{small}
\end{table*}

\begin{table*}
\begin{small}
\caption{Overview of NOEMA and 30m data used for SWAN }\label{tab:Dataquality}
\centering
\begin{tabular}{ll|ll|rr|rr|rr|rr}
\hline\hline
\noalign{\smallskip}
      & & & & \multicolumn{2}{c}{10 km\,s$^{-1}$} & \multicolumn{2}{c}{5 km\,s$^{-1}$} & \multicolumn{2}{c}{2.5 km\,s$^{-1}$} & \multicolumn{2}{c}{1 km\,s$^{-1}$} \\ 
    Line & $\nu_\mathrm{rest}$& native res. & b$_\mathrm{pa}$ & $T_\mathrm{peak}$ & rms  & $T_\mathrm{peak}$ & rms  & $T_\mathrm{peak}$& rms & $T_\mathrm{peak}$& rms \\
    Line & [GHz] & [$\arcsec\times\arcsec$] & [$^\circ$] & [mK]& [mK] & [mK] & [mK]  &  [mK]& [mK] & [mK]& [mK]\\
    (1) & (2) & (3) & (4) & (5) & (6) & (7) & (8) & (9) & (10) & (11) & (12)\\
\noalign{\smallskip}
\hline
\noalign{\smallskip}
       \multicolumn{12}{c}{NOEMA+30m data}  \\ 
\noalign{\smallskip}
\hline
\noalign{\smallskip}
\cch{}(1--0) & 87.3169 & 3.1$\times$2.7 & 49 & 141.0 & 12.4 & 121.4 & 15.7 & 204.7 & 22.9 & 308.1 & 33.4 \\
HNCO(4--3) & 87.9252 & 3.1$\times$2.7 & 49 & 197.0 & 8.4 &  - &  - &  - &  - &  - &  - \\
HCN(1--0) & 88.6319 & 3.0$\times$2.7 & 49 & 555.5 & 9.8 & 564.4 & 16.9 & 570.4 & 22.1 & 586.4 & 30.5 \\
\hcop{}(1--0) & 89.1885 & 3.0$\times$2.7 & 50 & 451.1 & 9.8 & 459.9 & 17.7 & 473.3 & 18.0 & 499.7 & 30.0 \\
HNC(1--0) & 90.6635 & 3.0$\times$2.6 & 49 & 308.8 & 9.6 & 343.6 & 13.5 & 361.5 & 18.8 & 379.1 & 27.8 \\
\nnhp{}(1--0) & 93.1738 & 2.9$\times$2.5 & 48 & 271.7 & 8.6 & 290.8 & 14.1 & 317.4 & 17.5 & 339.0 & 26.1 \\
C$^{18}$O(1--0) & 109.7822 & 2.4$\times$2.2 & 45 & 436.3 & 14.4 & 549.2 & 20.2 & 578.7 & 27.9 & 606.3 & 42.2 \\
HNCO(5--4) & 109.9058 & 2.4$\times$2.1 & 49 & 210.9 & 12.8 &  - &  - &  - &  - &  - &  - \\
$^{13}$CO(1--0) & 110.2014 & 2.4$\times$2.1 & 47 & 1461.1 & 16.3 & 1725.8 & 23.9 & 1763.6 & 31.0 & 1808.9 & 44.8 \\
\noalign{\smallskip}
\hline
\noalign{\smallskip}
       \multicolumn{12}{c}{NOEMA  data}  \\ 
\noalign{\smallskip}
\hline 
\noalign{\smallskip}
\cch{}(1--0) & 87.3169 & 3.0$\times$2.6 & 48 & 161.4 & 12.9 & 127.9 & 16.1 & 210.9 & 23.2 & 326.6 & 34.2 \\
HNCO(4--3) & 87.9252 & 3.0$\times$2.6 & 49 & 207.1 & 8.8 &  - &  - &  - &  - &  - &  - \\
HCN(1--0) & 88.6319 & 2.9$\times$2.6 & 47 & 570.8 & 10.3 & 577.5 & 18.0 & 584.2 & 22.8 & 596.9 & 31.8 \\
\hcop{}(1--0) & 89.1885 & 2.9$\times$2.6 & 47 & 448.2 & 10.2 & 469.3 & 18.9 & 481.1 & 17.7 & 511.6 & 31.1 \\
HNC(1--0) & 90.6635 & 2.9$\times$2.5 & 47 & 318.4 & 10.0 & 362.9 & 14.3 & 378.7 & 19.1 & 392.0 & 28.1 \\
\nnhp{}(1--0) & 93.1738 & 2.6$\times$2.6 & 88 & 283.2 & 9.6 & 290.2 & 14.9 & 323.2 & 18.1 & 341.6 & 26.1 \\
C$^{18}$O(1--0) & 109.7822 & 2.3$\times$2.1 & 47 & 448.5 & 15.0 & 550.7 & 21.1 & 577.3 & 28.5 & 599.5 & 43.5 \\
HNCO(5--4) & 109.9058 & 2.3$\times$2.1 & 47 & 219.2 & 12.6 &  - &  - &  - &  - &  - &  - \\
$^{13}$CO(1--0) & 110.2014 & 2.3$\times$2.1 & 47 & 1492.7 & 16.9 & 1755.2 & 24.6 & 1771.3 & 31.0 & 1802.6 & 44.0 \\

\noalign{\smallskip}
\hline
\noalign{\smallskip}
       \multicolumn{12}{c}{30m data} \\ 
\noalign{\smallskip}
\hline 
\noalign{\smallskip}
\cch{}(1--0) & 87.3169 & 29.7$\times$29.7 & 0 & 16.5 & 3.0 & 22.3 & 4.1 & 26.0 & 5.6 & 39.8 & 7.9 \\
HNCO(4--3) & 87.9252 & 29.5$\times$29.5 & 0 & 12.2 & 2.3 &  - &  - &  - &  - &  - &  - \\
HCN(1--0) & 88.6319 & 29.3$\times$29.3 & 0 & 56.4 & 2.3 & 58.8 & 3.3 & 62.6 & 4.6 & 67.8 & 6.2 \\
\hcop{}(1--0) & 89.1885 & 29.1$\times$29.1 & 0 & 36.6 & 2.5 & 39.2 & 3.4 & 44.8 & 4.6 & 51.8 & 6.4 \\
HNC(1--0) & 90.6635 & 28.6$\times$28.6 & 0 & 22.2 & 2.4 & 26.1 & 3.3 & 33.8 & 4.6 & 40.8 & 6.7 \\
\nnhp{}(1--0) & 93.1738 & 27.9$\times$27.9 & 0 & 17.2 & 2.0 & 22.8 & 2.9 & 27.5 & 4.1 & 31.4 & 6.1 \\
C$^{18}$O(1--0) & 109.7822 & 23.7$\times$23.7 & 0 & 44.2 & 3.8 & 50.9 & 5.0 & 60.2 & 7.0 & 72.0 & 9.9 \\
HNCO(5--4) & 109.9058 & 23.6$\times$23.6 & 0 & 15.8 & 3.7 &  - &  - &  - &  - &  - &  - \\
$^{13}$CO(1--0) & 110.2014 & 23.6$\times$23.6 & 0 & 154.9 & 3.9 & 170.4 & 5.6 & 178.4 & 7.6 & 190.1 & 11.5 \\
\noalign{\smallskip}
\hline
\noalign{\smallskip}
\end{tabular}
\tablefoot{Properties of SWAN data sets: For the combined NOEMA+30m data, the NOEMA data, and 30m data we list the molecular emission line (1), its rest frequency (2), native beam size (3) and beam position angle (4), as well as peak temperature $T_\mathrm{peak}$ and typical rms for different spectral resolutions (10\,km\,s$^{-1}$: (5),(6); 5\,km\,s$^{-1}$: (7),(8); 2.5\,km\,s$^{-1}$: (9),(10), 1\,km\,s$^{-1}$: (11),(12)). 
$T_\mathrm{peak}$ refers to the peak intensity of the brightest pixel in the data cube. The rms is the average rms of the first and last 5 channels, which are free of emission. 
For all data sets, both peak intensity and rms are calculated inside the area covered by the mosaic (see Figure~\ref{fig:mosaicpointings}). In this area, the sensitivity is comparably constant and to avoid the increased noise towards the edges of the mosaic.  The 30m data refers to combined data sets of archival EMPIRE and CLAWS data and the newly obtained IRAM 30m observations. }
\end{small}
\end{table*}

\subsection{IRAM-30m single-dish observations}

Spatially extended emission accounts for a significant fraction of the total flux in galaxies \citep[e.g., appendix~D in ][]{leroy_phangsalmapipeline_2021}. Since interferometers 
are not sensitive to emission from coarser scales,
we complement the NOEMA interferometric data with single dish data from the IRAM-30m telescope to provide the low spatial frequency (or ``short-spacing'') information. 

The SWAN 30m data partially consists of archival IRAM 30m observations. We used archival HCN(1--0), HNC(1--0) and \hcop{}(1--0) data from the IRAM-30m EMPIRE survey \citep{jimenez-donaire_empire_2019}, and \nnhp{}(1--0), \ceto{}(1--0) and \tco{}(1--0) data from the IRAM-30m CLAWS survey (055-17, PI: K. Sliwa; \citealt{den_brok_co_2022}). More detailed information about the EMPIRE and CLAWS observations can be found in the corresponding survey papers \citep{jimenez-donaire_empire_2019, den_brok_co_2022}. 
Both surveys cover a FoV that is significantly larger than that of the NOEMA observations. Within the NOEMA FoV, the EMPIRE and CLAWS observations represent about 14 hours of 30m observations (see also Table~\ref{tab:ObservationalSettings}), insufficient for our sensitivity requirements  and to avoid any degradation when combining 30m and NOEMA data (according to eq. 19 of memo IRAM-2008-2 \citep{rodriguez_shortspacings_2008}).  
We therefore obtained $\sim55$ hours of new observations with the IRAM-30m between 2020 February and April (project 238-19). M51 was observed with  EMIR combined with the Fast Fourier Transform Spectrometers (FTS). We used a frequency bandwidth of $2\times7.9$\,GHz, regularly sampled at 195\,kHz. We used on-the-fly (OTF) mapping mode with scan legs of lengths 200 and $300''$ along the RA and Dec axes, respectively. 
The distance between two scan legs (i.e., perpendicular to the scanning direction) was $8''$.
Additionally, we shifted the center of the mapped box in each iteration by multiples of 2\arcsec{} to get a final grid with a finer step of order 2\arcsec. 

We typically observed under good weather conditions (median PWV=$2^{+4}_{-2}$~mm, where, hereafter, the subscript and superscript indicate the offsets of the 16$^\mathrm{th}$ and 84$^\mathrm{th}$ percentiles from the median, respectively) and at high elevation ($69^{\circ\, {+10^\circ}}_{\phantom{\circ}\,{-16^\circ}}$).  
The typical system temperatures on the $T_A^*$ scale were $89^{+10}_{-12}$~K for lines below 91~GHz, $79^{+12}_{-6}$~K for N$_2$H$^+$(1-0), and $109^{+17}_{-16}$~K for the CO isotopologue lines and HNCO(5-4).  

The absolute flux calibration of the 30m observations was monitored with the bright carbon-rich AGB star IRC+10216, which was observed for a few minutes at the start of most of the observing runs. For each observation, we extracted and integrated ten well-detected lines (signal-to-noise ratio SNR $\gtrsim50$) with frequencies spaced across the observed bandwidth ($88.6-110.2$~GHz). This allowed us to check the relative calibration of the telescope as a function of time and frequency disregarding thermal noise effects. 
We confirmed that the spectral shape of every line is constant with time up to a varying scaling factor. 
This suggests that the observed amplitude variations are not caused by pointing errors. 
In short, since the lines are distributed in different ways across the envelope of IRC+10216, pointing errors would tend to change the shape of the lines with widespread emission. 
We also found that the correlation between the integrated intensities of pairs of lines was better when considering lines within the same EMIR subband, suggesting that the observed amplitude variations are driven by calibration errors that depend sensitively on frequency.  
Overall, the rms uncertainty in the integrated intensities is of order $\sim5\%$ for most molecular lines studied, and $\sim 10-15\%$ for the four lines in the 93-102\,GHz regime. 
We also detected systematic differences between polarizations of order $\sim5-10\%$ for all lines. 
In the end, we can assume a relative flux calibration uncertainty on the order of $5-10\%$ for all lines, consistent with typical expectations.

\section{Data reduction and imaging}
\label{sec:Datareduction}

\subsection{NOEMA data reduction and imaging }
\label{sec:NOEMAonlydata}

We calibrated the NOEMA observations using the standard \texttt{GILDAS/CLIC}\footnote{\url{https://www.iram.fr/IRAMFR/GILDAS}} pipeline. We extracted calibrated $uv$ tables for each velocity resolution (i.e., 10, 5, 2.5, and 1\,km\,s$^{-1}$) and line before imaging with \texttt{GILDAS/MAPPING}. The exact spectral extent of the resulting cubes depends on the line, as described in Section~\ref{sec:NOEMAobs}. 
We did not subtract a continuum in the visibilities in order to avoid biasing the very broad line (typically several hundred\,km\,s$^{-1}$) that appears near the galaxy center. 
The produced cubes thus contain a contribution from continuum emission. 
However, for the purposes of line analysis, this continuum contribution is only significant in 
very small regions along the southern nuclear radio jet axis 
\citep[XNC;][]{ford_bubbles_1985}. 
Hereafter, we subtract this contribution in the image plane when needed (see Section~\ref{sec:NOEMA30mdata}).
We imaged all the $uv$ tables on the same spatial grid, which we centered at RA=13:29:52.532, Dec=47:11:41.982 (J2000). The grid has a pixel size of 0.31\arcsec{} and a total map size of $768\times1024$ pixels. 
Cleaning was performed with the Högbom cleaning algorithm with a constant number of clean components. The number of clean components was chosen to obtain residuals that look like noise (i.e., no coherent spatial structures are present any longer).
Since the emission of the brighter lines (e.g., \tco) extends across the full FoV we did not use any support (i.e., cleaning mask) during the cleaning. We used 64\,000 clean components for the 10 and 5\,km\,s$^{-1}$ cubes and 32\,000 for the 2.5 and 1\,km\,s$^{-1}$ ones. 
We used fewer clean components for the higher spectral resolution cubes since the signal-to-noise is lower at higher spectral resolution. 
Finally, we converted the intensity scale from Jy/beam to Kelvin with the standard \texttt{GILDAS-MAPPING} GO JY2K command. 

The final NOEMA data set yields detections of nine molecular lines between $\sim$87.1 and 110.3 GHz: 
C$_2$H(1--0), HNCO(4--3), HCN(1--0),  HCO$^+$(1--0), HNC(1--0),\nnhp{}(1--0), C$^{18}$O(1--0), HNCO(5--4) and $^{13}$CO(1--0).
We detect the two brightest \cch{}(1-0) hyperfine transitions at rest-frequencies of 87.317 and 87.402 GHz, as well as a fainter transition at 87.284 GHz overlapping along the velocity axis of the first transition mentioned.
We do not attempt to separate them, but instead generate one data cube that contains all the detected \cch emission. 
The spectral axis of the delivered \cch data cube is centered on 87.3169 GHz (N= 1-0, J=3/2-1/2, F= 2-1).
We list the rest frequencies of our detected lines in Table~\ref{tab:Dataquality}, as well as 
the typical RMS, peak temperature and native resolution of the data cubes with 10, 5, 2.5, and 1\,km\,s$^{-1}$ spectral resolution.

\subsection{30m data reduction and imaging }
\label{sec:30mdatareduction}

The new IRAM 30m data were reduced to a set of common spectral and spatial grids to ensure a homogeneous treatment for all lines. 
First, the data for each target line was isolated by extracting frequency windows of 350\,MHz ($\sim 950-1200$\,km\,s$^{-1}$) centered on the rest frequency of each line. For \cch{}(1-0), we increased the window width to 400\,MHz to ensure that all the hyperfine structure components were included. 
Using the Ruze formula with the \texttt{CLASS} command \texttt{MODIFY BEAM\_EFF /RUZE}, the temperature scale was converted from $T_A^{\star}$ to $T_\mathrm{mb}$. 
Next, the data were spatially reprojected to match the NOEMA projection center of RA=13:29:52.532, Dec=47:11:41.982 (J2000). 
The Doppler correction was then recomputed for each spectrum to take into account 1) the change of velocity convention from optical during the observations to radio during the analysis, and 2) its variation as a function of position (because the Doppler tracking is computed only at the projection center and kept fixed during each scan to avoid creating standing waves). The velocity scale was updated to ensure that the redshifted frequency of the line in the local standard of rest (LSR) frame corresponds to a systemic velocity of $0\,$km\,s$^{-1}$. 
The mean RMS noise across each data cube was measured using the baseline residuals. 
We removed spectra for which the RMS noise was greater than three times the mean value in the cube ($\sim3$\% of all acquired spectra were rejected in this step). 

The IRAM-30m observations were imaged using standard \texttt{GILDAS/CLASS} procedures. The data was regridded spectrally to match the NOEMA data, with four different spectral resolutions: 10\,km\,s$^{-1}$, 5\,km\,s$^{-1}$, 2.5\,km\,s$^{-1}$ and 1\,km\,s$^{-1}$. All  were imaged using a Gaussian kernel of FWHM $\sim 1/3$ the 30m HPBW.  
We produced two sets of 30m data products with different spatial grids. The first grid has a pixel size of $4''$ for distribution as a stand-alone product. The second grid is identical to the one used to grid the NOEMA data to ensure a good spatial sampling when merging the single-dish and interferometric datasets.
We list the typical peak intensities and noise levels of the SWAN 30m data in Table~\ref{tab:Dataquality}. 

We conducted several checks on the 30m data reduction and imaging, which we describe in more detail in the appendices. Specifically, we tested for consistency between the new and archival 30m data sets (Appendix~\ref{app:archival_vs_new30m}). 
Overall, we find that the data sets agree to within 10\%. 
Finally, we analyse the impact of the 30m error beams onto the flux filtered out by the interferometer in Appendix~\ref{app:errorbeam}.  For \tco, the contribution of the error beam is ${<}10\%$ for the vast majority (i.e. 98\%) of sightlines, and the median error beam contribution is 1.5\% per pixel inside the NOEMA SWAN FoV. 
As the main beam efficiency decreases with increasing frequency, we expect the error beam contribution to be less significant for other emission lines in the SWAN survey.

\subsection{Combined NOEMA+30m imaging}
\label{sec:NOEMA30mdata}

The 30m data is combined with the NOEMA data in the $uv$ plane using the \texttt{GILDAS/MAPPING} \texttt{UV\_SHORT} command \citep[see][for details]{pety_revisiting_2010}. 
For the combination, we use the calibrated but non-continuum subtracted NOEMA $uv$ tables and the baseline-subtracted single-dish $uv$ tables. This is slightly inconsistent,
but there is currently no reliable method to measure the continuum emission at 3\,mm with the IRAM 30m. 
For several lines, this can result 
in the noise level in the center being on average offset towards slightly positive values
(compare Figure~\ref{fig:Spectra}).
For \nnhp{}(1-0) and HNCO(4-3), we apply an order 1 baseline subtraction in the final data cube during post-processing. 
We confirm that the baseline subtraction
does not affect any of the quantitative results for these lines. 
To verify this, we performed all analyses shown below using both the baseline-corrected and uncorrected data cubes for \nnhp.  The overall scientific conclusions and observed trends remain unchanged.

We image the combined data in a similar way as for the NOEMA-only data (Section~\ref{sec:NOEMAonlydata}). The resulting data cubes have the same spatial grid as for the stand-alone NOEMA data (i.e.,  a pixel size of 0.31\arcsec, which is $\sim 7-10$ times smaller than the native resolution beam size, on a grid of 768$\times$1024 pixels) and cleaned via Högbom-cleaning without cleaning masks until the cleaned flux reaches a 
stable number. 
We used twice as many clean components as for the NOEMA-only data (128\,000 clean components for the 10 and 5\,km\,s$^{-1}$ cubes and 64\,000 for the 2.5 and 1\,km\,s$^{-1}$ ones), since the short-spacing data introduces additional complexity.
Lastly, we convert the intensity scale from Jy/beam to Kelvin. 
We confirmed that the noise is "well-behaved" in each NOEMA+30m line cubes. The rms noise level shows little dependence on frequency across the whole bandwidth, if at all, and there is negligible correlation between adjacent channels.

\subsection{Flux recovery}
\label{sec:Fluxrecovery}

In Table~\ref{tab:FluxRecovery}, we list the fraction of flux recovered by the native-resolution NOEMA observations for different spectral versions of the SWAN data cube for each emission line. The flux recovery is calculated by smoothing and spatially regridding the NOEMA data to match the 30m data (the spectral grid of the NOEMA data is already matched to the 30m data during the imaging), then summing the emission from the central $100\times100\arcsec$ across all channels.
The ratio of integrated NOEMA flux to 30m flux yields the flux recovery estimate.
We limit the FoV during this computation to avoid contribution from the increased noise toward the edges of the mosaic (Figure~\ref{fig:mosaicpointings}).
For most lines, the NOEMA observations recover $\lesssim50\%$ of the 30m flux at 10\,km\,s$^{-1}$ spectral resolution. This fraction is consistently lower for all lines when the data products are imaged with narrower channels, falling to a typical value of $20-30\%$ flux recovery at a spectral resolution of 1\,km\,s$^{-1}$.  
This is probably due to the lower sensitivity of the data at higher spectral resolution, which affects the deconvolution.
Our method for calculating flux recovery is not effective for the fainter lines like \nnhp, \cch, and HNCO. 
Regridding and smoothing the NOEMA-only data of those lines to match the 30m data strongly reduces the line SNR.
As a result, we cannot make precise conclusions about these specific lines (uncertainties as large as $\gtrsim$80\% of the flux recovery value). Table~\ref{tab:FluxRecovery} only lists the flux recovery for lines with peak temperature to RMS ratio of $\gtrsim$10 at a spectral resolution of 10\,km\,s$^{-1}$ in the 30m-only data (compare Table~\ref{tab:Dataquality}). 

\begin{figure}
     \centering
     \includegraphics[width = 0.5\textwidth]{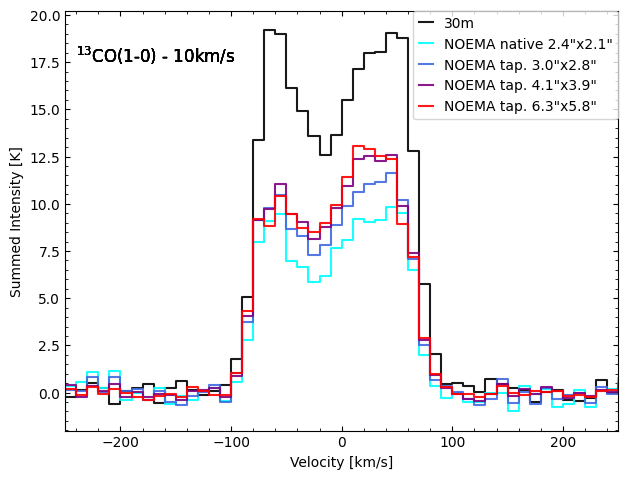}
     \caption{Flux recovery for \tco{}(1-0) at 10\,km\,s$^{-1}$ spectral resolution. The native resolution NOEMA data (cyan line) is convolved and regridded to match the 30m data (black line) in resolution. The flux shown is the summed flux per channel inside the central $100\times100\arcsec$.  NOEMA data tapered to coarser spatial resolutions of about 3, 4, and 6$\arcsec$ (blue, purple, red lines) have higher recovery rates. }
     \label{fig:Fluxrecovery_Tapering}
 \end{figure}

We also studied the relation between synthesized angular resolution and flux recovery. We tapered the data during the data reduction to coarser angular resolutions of 3, 4, and 6$\arcsec$ and recalculated the flux recovery per channel. 
Figure~\ref{fig:Fluxrecovery_Tapering} shows the flux recovery of the native resolution NOEMA $^{13}$CO(1--0) data compared to the 30m data for each channel as a function of the angular resolution associated to different tapering distance in the $uv$ plane.
At fixed spectral resolution, the flux recovery of the NOEMA data improves as the angular resolution increases from 2 to 4$\arcsec$.  
Above $\gtrsim 4\arcsec$, the flux recovery converges and no longer increases with increasing beam size. 
Our SWAN flux recovery results are consistent with findings for the PAWS survey \citep{schinnerer_pdbi_2013}, where only a marginal improvement in the flux recovery was reported when the resolution was degraded from 3\arcsec{}to 6\arcsec{}\citep{pety_plateau_2013}.  Our tests indicate that all the flux present in the SWAN NOEMA data is deconvolved at scales $\gtrsim 4\arcsec$.

\begin{table}
\begin{small}
\caption{Interferometric flux recovery fraction}\label{tab:FluxRecovery}
\centering
\tabcolsep=0.11cm
\begin{tabular}{l|llll}
\hline\hline
\noalign{\smallskip}
    Line & \multicolumn{4}{c}{Flux recovery }\\
\noalign{\smallskip}
 & 10\,km\,s$^{-1}$ &  5 km\,s$^{-1}$ & 2.5 km\,s$^{-1}$ & 1 km\,s$^{-1}$\\
\hline
\noalign{\smallskip}
\hline
$^{13}$CO(1-0) & $0.48\pm0.01$ & $0.45\pm0.01$ & $0.42\pm0.01$ & $0.33\pm0.01$ \\
C$^{18}$O(1-0) & $0.29\pm0.04$ & $0.18\pm0.04$ & $0.23\pm0.04$ & $0.15\pm0.02$ \\
HCN(1-0) & $0.43\pm0.02$ & $0.40\pm0.04$ & $0.33\pm0.02$ & $0.25\pm0.01$ \\
HCO$^+$(1-0) & $0.42\pm0.06$ & $0.39\pm0.04$ & $0.32\pm0.02$ & $0.23\pm0.03$ \\
HNC(1-0) & $0.49\pm0.07$ & $0.40\pm0.06$ & $0.35\pm0.05$ & $0.23\pm0.04$ \\
    \hline 
    \noalign{\smallskip}

\end{tabular}
\tablefoot{The data is extracted from the central $100\times100\arcsec$ of the SWAN FoV. The native angular resolution NOEMA data is smoothed and regridded to match the 30m data spatially and spectrally. The tabulated values are the summed flux measured from the NOEMA data, divided by the summed flux measured from the 30m data. The quoted errors are statistical uncertainties only. We tabulate measurements for data imaged at spectral resolutions of 10, 5, 2.5, and 1 km\,s$^{-1}$. 
Each lines native angular resolution is the same across all spectral resolutions (see Table~\ref{tab:Dataquality}). 
}
\end{small}
\end{table}

\subsection{Signal in the NOEMA+30m data cubes}
\label{sec:SignalinN30}

\begin{figure*}
    \centering
    \includegraphics[width = 0.33\textwidth]{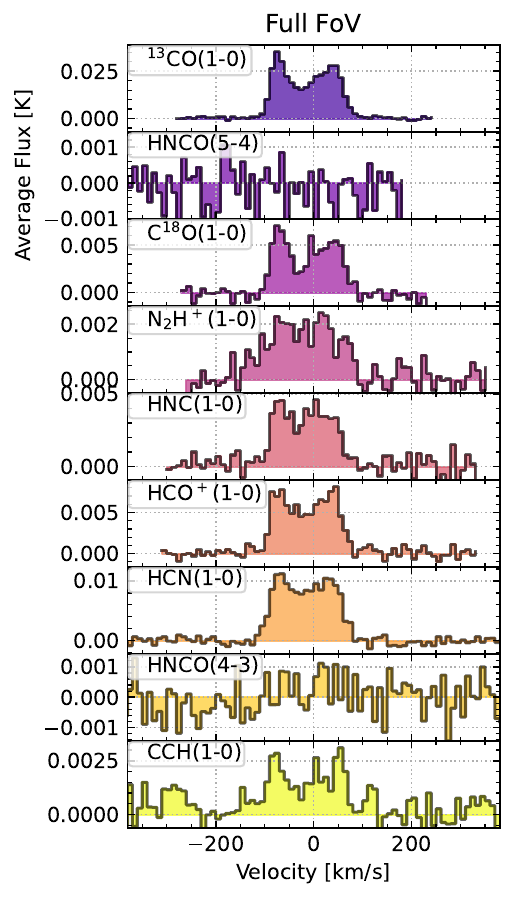}
    \includegraphics[width = 0.31\textwidth]{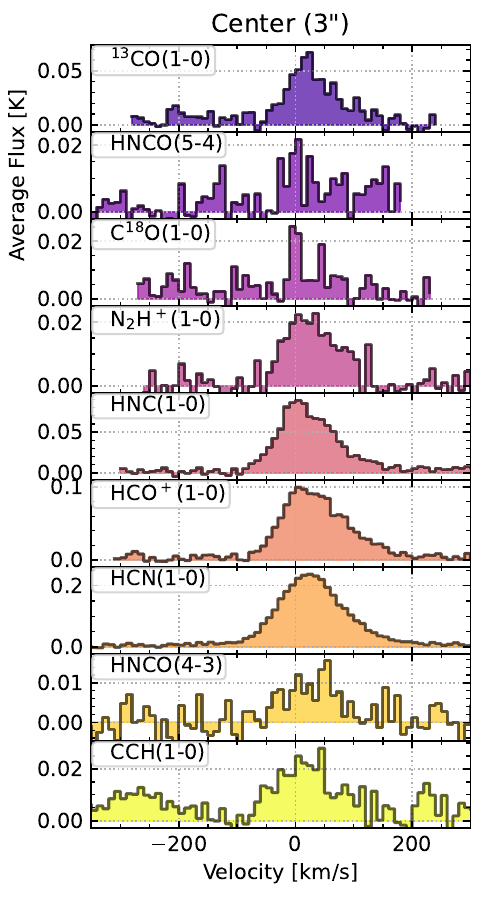}
    \includegraphics[width = 0.31\textwidth]{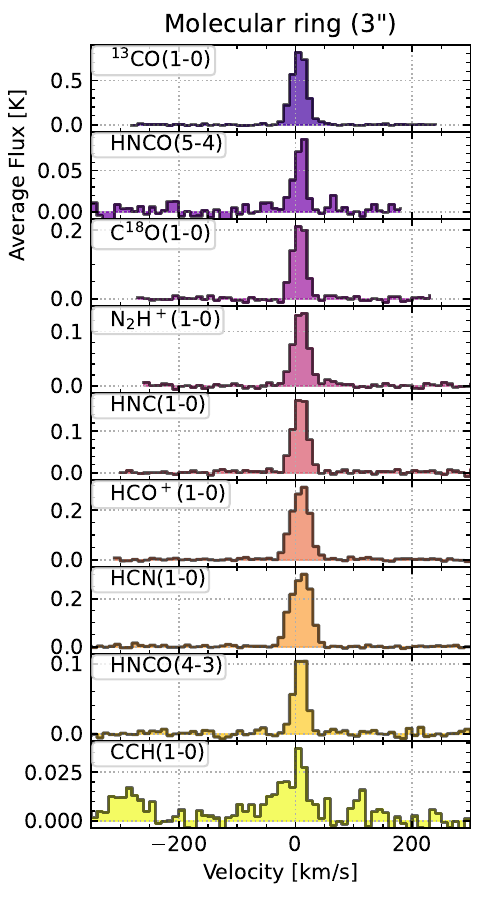}
    \caption{Spectra of all molecular lines for the full disk (left panel), a 3\arcsec{}sized region in radius in the center (middle panel) and the western edge of the molecular ring (right panel) where \nnhp is particularly bright. 
    The full disk spectra are extracted using the area covered by the perimeter (``hull'') of the mosaic of all pointings (see Figure~\ref{fig:mosaicpointings}) for \tco. Since this hull is frequency dependent, this corresponds to the smallest area out of all lines.}
    \label{fig:Spectra}
\end{figure*}

The NOEMA+30m data is imaged at spectral resolutions of 10, 5, 2 and 1\,km\,s$^{-1}$. 
We list the typical RMS and peak temperature per channel for each spectral resolution at the native angular resolution for all lines in Table~\ref{tab:Dataquality}. 

Figure~\ref{fig:Spectra} shows spectra at 10\,km\,s$^{-1}$ spectral resolution for the combined NOEMA+30m data for all detected molecular lines: \tco{}(1-0), \ceto{}(1-0), HNCO(5-4), \nnhp{}(1-0), \hcop{}(1-0), HNC(1-0), HCN(1-0), HNCO(4-3), and \cch{}(1-0). 
We show spectra of the full FoV inside the perimeter (hereafter ``hull'') covered by the mosaics (compare Figure~\ref{fig:mosaicpointings}), as well as inside a 3\arcsec{} sized region located at the galaxy center and the bright spot at the south-western edge of the molecular ring (RA 13:29:50.0633, DEC: 47:11:25.2040 (J2000)). 
As expected, most central spectra, especially from HCN, HNC and \hcop, are broader than those in the molecular ring, likely due to the complex kinematics due to the AGN-driven outflow in the galaxy center. 
The multiple peaks in the \cch{}(1-0) spectra are hyperfine transitions.

\begin{figure*} 
    \centering
    \includegraphics[width = \textwidth]{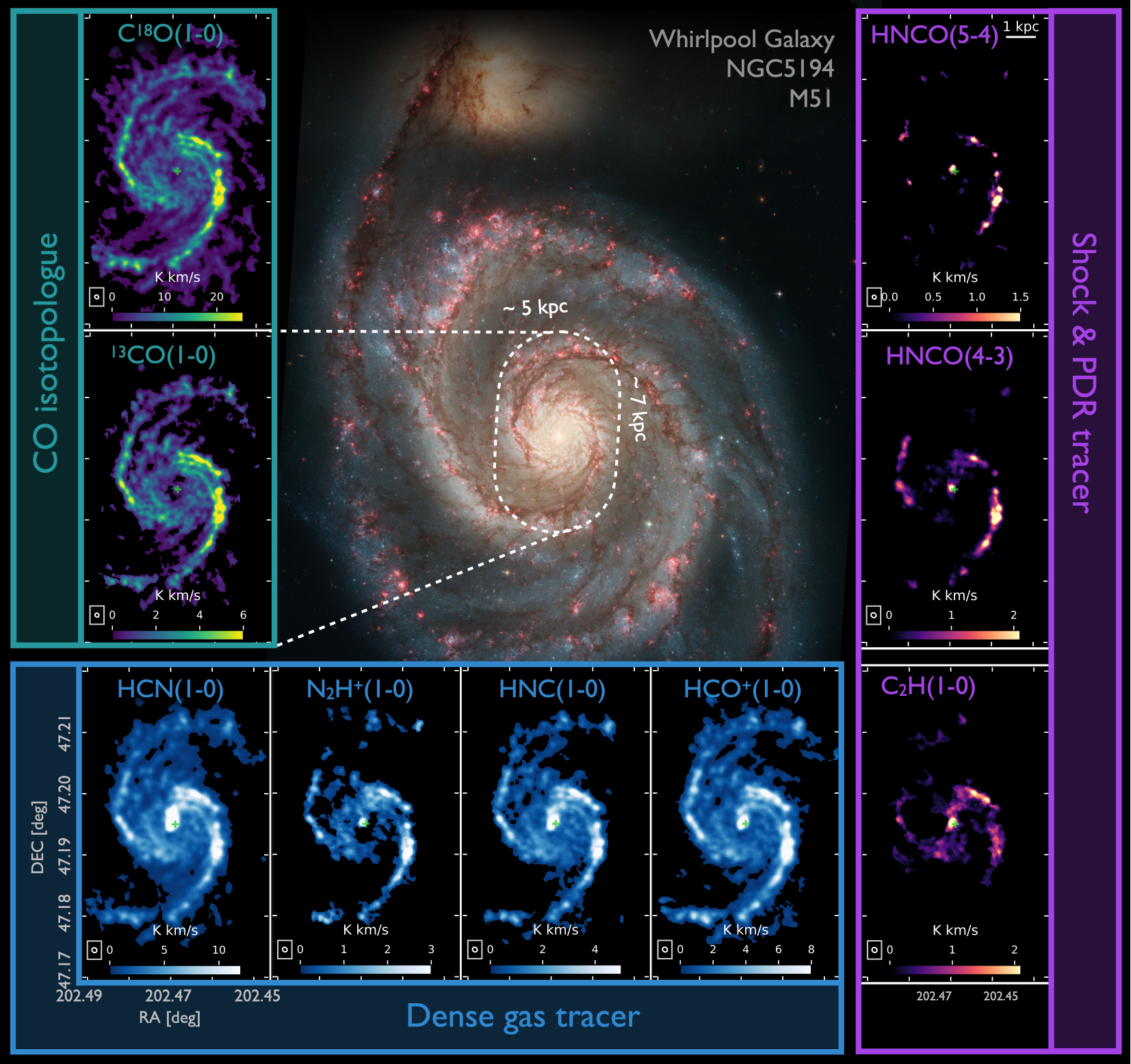}
    \caption{Integrated intensity (moment-0) maps of the SWAN data set (combined NOEMA and IRAM 30m observations) for the J=1--0 transitions of \tco, \ceto, \nnhp, \hcop, HNC, HCN, and \cch{} plus HNCO(J=4--3) and HNCO(J=5--4) at their native angular resolution ($\sim2.3-3.1\arcsec$). The lines are grouped in different subsets based on their commonly used applications.
    The maps are created with the \texttt{GILDAS} ``Island-method'' (see Appendix~\ref{app:Momentmappipelines}).  
    We show the beam size in the bottom left of all panels as well as a 1 kpc scale bar in the top right panel. 
    The outline of the SWAN FoV is indicated by a dashed ellipse on top of a multi-color HST image (credit: S. Beckwith (STScI) Hubble Heritage Team, (STScI/AURA), ESA, NASA). 
    }
    \label{fig:Gallery_overview}
\end{figure*}

\subsection{Moment map creation}
\label{sec:m0mapcreation}

The final SWAN data cubes of combined NOEMA and IRAM-30m data are publicly available\footnote{\url{https://oms.iram.fr/dms} and \url{https://www.canfar.net/storage/list/}}.
Integrating all emission within a fixed velocity range at all lines of sight will add noise to the already faint emission. Therefore, a number of masking techniques that select regions and channels within the data cube to integrate emission have been used in the literature and depending on the method used the total recovered flux this will differ \citep[see, e.g., Appendix B in ][]{pety_plateau_2013}.
The advantages and disadvantages of different masking strategies depend on the science objective, and the relative importance of completeness versus avoiding false positives \citep[see, e.g.,][]{leroy_phangsalmapipeline_2021}
Utilizing these data cubes at a spectral resolution of 10\,km\,s$^{-1}$, we test two commonly used methods of moment-map creation. 

First, we create moment-maps with the \texttt{GILDAS/CUBE} ``Island-method''~\citep{einig_deep_2023}. The resulting maps are shown in Figure~\ref{fig:Gallery_overview} (and for a better comparison with an alternative method in Figure~\ref{fig:Gallery_GILDAS} in Appendix~\ref{app:Momentmappipelines}) at each line's native angular resolution. 
For each line, this method identifies connected structures as follows: 
We calculate the noise in channels with velocities $|v| >200\,$km\,s$^{-1}$. 
We then smooth the data cube with a Gaussian kernel of size $B_{maj}\times B_{min}*PA$ and calculate the signal to noise ratio (SNR) at each pixel based on the smoothed cube and the calculated noise. 
Next, we identify connected structures above a selected threshold of SNR = 2 in the position-position-velocity (ppv) cube. 
These structures are applied to the original data cube and emission is integrated over the pixels within those structures. 
By doing this for each line at its native resolution, we minimize noise that would otherwise be added in the integration process, while still conserving fainter emission from connected structures.  
Figure~\ref{fig:Gallery_overview} (and Figure~\ref{fig:Gallery_GILDAS}) show the moment-0 maps created when selecting structures based on each line individually at their native angular resolution.  
In addition to moment-0 maps, we create velocity field maps (moment-1), line-width maps (moment-2), peak temperature maps as well as associated uncertainty maps, which are available in the public data release. 

In contrast to selecting connected significant emission structures identified in the same data cube that is being used to generate moment maps, another common strategy is to construct a significant emission mask based on a single bright line (e.g., \CO, aka the ``prior'') and applying the resulting mask to the datacubes of fainter lines. 
Since M51 has an unusual center that is mostly devoid of \CO emission (including the \tco and \ceto isotopologues, see Figures~\ref{fig:Spectra} and~\ref{fig:Gallery_overview}), a \CO-based prior does not accurately capture the emission from all the lines in the SWAN survey field.
Bright HCN HNC, and \hcop emission in the center of M51, for example, are evident in Figure~\ref{fig:Spectra}.
The spectra inside a 3\arcsec{} sized central aperture (Figure~\ref{fig:Spectra}) are comparably broad for all three of these lines, with FWHM $\sim100\,$km\,s$^{-1}$.
We generate another set of moment maps using both \CO and HCN emission to construct a mask. This mask contains regions in which either HCN or \CO is detected. 
This mask captures the bulk molecular gas distribution best traced by \CO emission outside of the central kiloparsec, as well as the central region, which is best traced by the bright HCN emission (Figure~\ref{fig:Gallery_overview}).  
This is done by using the so-called ``PyStructure''\footnote{Pystructure documentation: \url{https://pystructure.readthedocs.io/en/latest/}} code \citep{den_brok_co_2022,neumann_almond_2023} and after convolving the data to a common resolution of 3.05\arcsec. 
High-resolution \CO(1-0) data is taken from the PdBI Arcsecond Whirlpool Survey \citep[PAWS;][]{schinnerer_pdbi_2013} and matched to our resolution.
Pixels in ppv space are thus selected for integration if either HCN or \CO is detected.  
The PyStructure code further allows the user to re-sample the data with hexagonal pixels, which capture the circular beam of the observations well. 
We hexagonally resample all data to a matched grid with 2 hexagons across each beam length. Integrated moment maps are then created by selecting the spectral windows where both \CO and HCN are significantly detected.  
This data is saved in a \texttt{numpy} table, we'll refer to as a ``PyStructure table'' from hereon.

While the former method is best suited when investigating individual lines, the latter is 
often preferable for the comparison of several lines, since it ensures that the same pixels of the ppv cube are used for integration. 
However, it tends to increase the noise in the integrated emission maps of fainter lines (e.g., HNCO(5-4), HNCO(4-3) and \cch{}(1-0)).
This is apparent when we compare Figure~\ref{fig:Gallery_GILDAS} and \ref{fig:Gallery}, where differences between them are driven by either the difference in methodology, or the difference in resolution (and therefore SNR).
As an example, HNCO(5-4) is shown at its native 2.3\arcsec{} in Figure~\ref{fig:Gallery_GILDAS}, but smoothed to a 3\arcsec{} resolution in \ref{fig:Gallery}. 
As the native resolution of \cch is $\sim3\arcsec$, differences in the moment maps arise due to the differences in methodology. 

Given both the \texttt{GILDAS} and PyStructure methods are useful for different analysis, we conduct a pixel-by-pixel comparison for \tco in Appendix~\ref{app:Momentmappipelines}. 
Overall, we find good agreement between both methods, with the \texttt{GILDAS} method recovering more flux than the Pystructure at lower intensities. 

The SWAN public data release includes data cubes and the moment maps created with \texttt{GILDAS} and are available on the IRAM Data Management System and the Canadian Advanced Network for Astronomical Research (CANFAR). 
The PyStructure table is utilized in upcoming SWAN publications analyzing the CO isotopologues \citep{denBrok2025, Galic_submitted} as well as dense gas tracers HCN, HNC, \hcop and \nnhp (Stuber et al. in prep).
The PyStructure table are distributed as a complementary product on the CANFAR.

\section{Results: 3mm line emission across M51}
\label{sec:Results}

We present an overview of the integrated intensity maps (moment-0) of all lines from the SWAN survey in Figure~\ref{fig:Gallery_overview}. The resulting maps from the two different methods of moment map generations (see Section~\ref{sec:m0mapcreation}) can be compared via Figs. \ref{fig:Gallery_GILDAS} and \ref{fig:Gallery}).
To compare the emission of all detected molecular lines, we utilize the moment maps created based on common priors (\CO and HCN, Figure~\ref{fig:Gallery}), created with the PyStructure-code. 

In this Section, we compare the spatial distribution of the integrated molecular line emission across the full FoV below. 
Further, we compare emission of molecular lines against each other (Section~\ref{sec:Science:Linevsline}), and compare line ratios with \CO between the central 1\,kpc and the disk (Section~\ref{sec:Science:LinevsCO}).  
In Section~\ref{sec:Science:Literature}, we compare the SWAN data set to high- and low-resolution observations from the literature for individual lines.

To first order, the SWAN maps show that 3mm molecular line emission is similarly distributed across M51's inner disk. For all tracers, the emission is bright along the spiral arms and in the molecular ring. Roughly half of the SWAN emission lines are also bright in M51's center, where an AGN with low-inclined radio jet is located. The notable exceptions are \tco and \ceto, which -- similar to \CO \citep[PAWS;][]{schinnerer_pdbi_2013} -- show relatively faint emission in the central region compared to elsewhere in the inner disk. 

Within the disk, the emission of all lines is particularly bright along the western side of the molecular ring, at the base of the southern spiral arm. 
Some lines, including \nnhp{}(1-0), \ceto{}(1-0), \tco{}(1-0) and HNCO(4-3) are brightest in this particular region (RA: 13:29:50.06332, DEC: 47:11:25.20404 (J2000)). 
Spectra of all lines inside a 3\arcsec-aperture centered on this region and centered on the galaxy center (Figure~\ref{fig:Spectra}) reveal that most lines reach higher intensities in this region compared to the center.
The \nnhp emission in this region was previously studied by S23, who reported unusually high \nnhp-to-HCN and \nnhp-to-\CO ratios compared to elsewhere in M51's inner disk. Other lines in the SWAN data set, such as the shock-tracer HNCO, are also bright in this region. This region will be investigated in more detail, using the full suite of SWAN emission lines, in a forthcoming paper (Stuber et al., in prep.).

\subsection{Comparison of line intensities}
\label{sec:Science:Linevsline}

\begin{figure*}
    \centering
    \includegraphics[width = 1\textwidth]{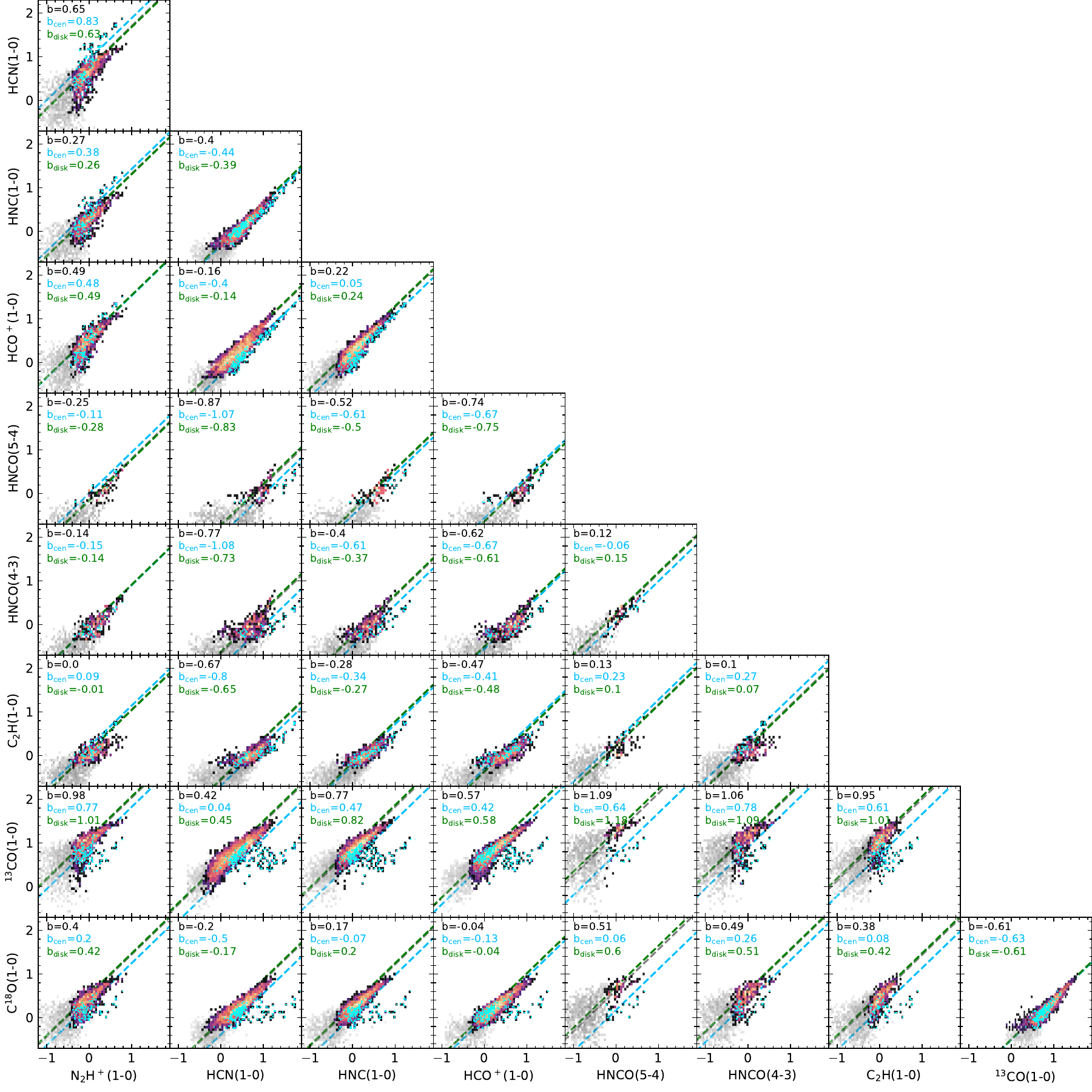}
    \caption{Logarithmic integrated line emission in K km\,s$^{-1}$ of all lines compared on a pixel-by-pixel scale. We show the 2D distribution of emission from both significant detections (colored points, $>$3$\sigma$) and non-detections (grey points, $<3\sigma$). The color scale of both detected and non-detected points indicates the point density and is for visual purpose only.
    We mark pixels inside the central 1\,kpc (in diameter, $\sim8\arcsec$) in cyan. 
    The grey dashed line corresponds to a power-law with a slope of unity and offset $b$ being the average line ratio calculated from pixels with significant detections, including the central pixels. We define $b = \mathrm{mean} \left(\mathrm{log}_{10} (y/x) \right)$, with $x$ and $y$ referring to the values on the x and y-axis. 
    We further show $b_\mathrm{cen}$ (blue dashed line) and $b_\mathrm{disk}$ (green dashed line) which are calculated using only pixels inside and outside the central 1\,kpc, respectively. $b_\mathrm{disk}$ and $b$ often overlap.
    Uncertainties are calculated following Gaussian error propagation and are listed together with all values of $b$ in Table~\ref{tab:offsetb} in Appendix~\ref{app:Offsets}.}
    \label{fig:Lineemissioncomparison}
\end{figure*}

In Figure~\ref{fig:Lineemissioncomparison}, we present a pixel-by-pixel comparison of the integrated intensity for all emission line pairs in the SWAN dataset.  
We highlight data points inside the central, inclination-corrected 1\,kpc (in diameter) region, and overlay a linear relation through the average line ratio to aid visual inspection. 
As shown in S23, an aperture of 1\,kpc captures most of emission that is likely affected by the AGN, and is in agreement with the spatial extent of optical AGN-typical line ratios, X-ray and radio emission \citep{blanc_spatially_2009}. 
We note that within this area, a nuclear bar coexists.
This central region contains $\sim200$ hexagonal pixels.
For each panel, we calculate the mean logarithmic line ratio ($b = \mathrm{mean} \left(\mathrm{log}_{10} (y/x) \right)$, with $x$ and $y$ referring to the values on the x and y-axis) for pixels where both of the lines involved in the line ratio are significantly detected ($>3\sigma$). All values are listed in Table~\ref{tab:offsetb} in Appendix~\ref{app:Offsets}. 
Additionally, we list $b_\mathrm{cen}$ and $b_\mathrm{disk}$, which refer to the same calculations performed inside and outside the central 1\,kpc, respectively. 
We add the significance (see Section~\ref{app:Offsets}) of the difference of $b_\mathrm{cen}$ and $b_\mathrm{disk}$ in Table~\ref{tab:offsetb}. 
Only a few pixels in the SWAN FoV show significant HNCO(5-4) detections, so the regression results including this line are highly uncertain. The few detections of this line are located in the center of the galaxy, as well as the south-western edge of the molecular ring, where all lines are remarkably bright. \\
We find the following: 
\begin{enumerate}
    \item[(a)] Most emission line pairs exhibit a roughly linear (slope of 1) correlation, when pixels in the central 1\,kpc are excluded. 
    This is particularly the case for emission 
    lines that are associated with denser molecular gas, such as  HCN, HNC and \hcop,
    which visually follow a linear correlation over 1-2 orders of magnitude. 
    The isotopologues \ceto and \tco are both also visually well-correlated with HCN, HNC, and \hcop.  

    \item[(b)] In most panels, pixels in the central region show a clear offset relative to pixels elsewhere in the disk, indicating that the line emission is driven by different mechanisms in the disk compared to the center.  
    HCN, HNC, \hcop, \nnhp, HNCO, and \cch emission in the center are significantly enhanced compared to both \ceto and \tco emission (compare Table~\ref{tab:offsetb}).  
    HCN emission in the center is clearly enhanced compared to all other lines. The enhancement is strongest when compared to \tco, and weakest relative to its isomer, HNC.
    Line combinations that do not vary significantly between the center and disk regions are \ceto and \tco, \nnhp and \hcop, and combinations including the fainter HNCO and \cch lines. \nnhp{}(1-0) is enhanced in the center compared to \ceto and \tco, but it is fainter compared to HCN, HNC and \hcop. 

    \item[(c)] Correlations in the disk that visually appear to deviate from a linear trend include the trends between \nnhp and most lines except HNCO(4-3), and between \cch and most  lines except \tco and potentially \ceto.      
    We note that these visual trends are mostly driven by the brightest pixels of each line and could be biased by the differences in SNR. 
    As an example, in S23 we measured a super-linear trend between \nnhp and HCN emission ($m=1.2$), which was mainly driven by the brightest pixels in \nnhp emission. 
    
\end{enumerate}

Our data show that while all molecular lines are bright in similar regions in the disk (Figure~\ref{fig:Gallery_overview}), there are significant variations in the line ratios, both on 125\,pc scales and across larger $\sim$kpc-scale environments such as the center. 
Moreover, the not exactly linear correlation between several molecular emission lines 
suggests that these cloud-scale observations are sufficient to 
detect some variations in the excitation, chemical abundance, and opacity in the molecular gas across M51's disk.
Overall, the variations in line emission ratios between the central 1\,kpc, which potentially arise due to the AGN, and the disk
are larger than the variations observed elsewhere in the  disk.

\subsection{Global line ratios with \CO}
\label{sec:Science:LinevsCO}

\begin{figure} 
    \centering
    \includegraphics[width = 0.48\textwidth]{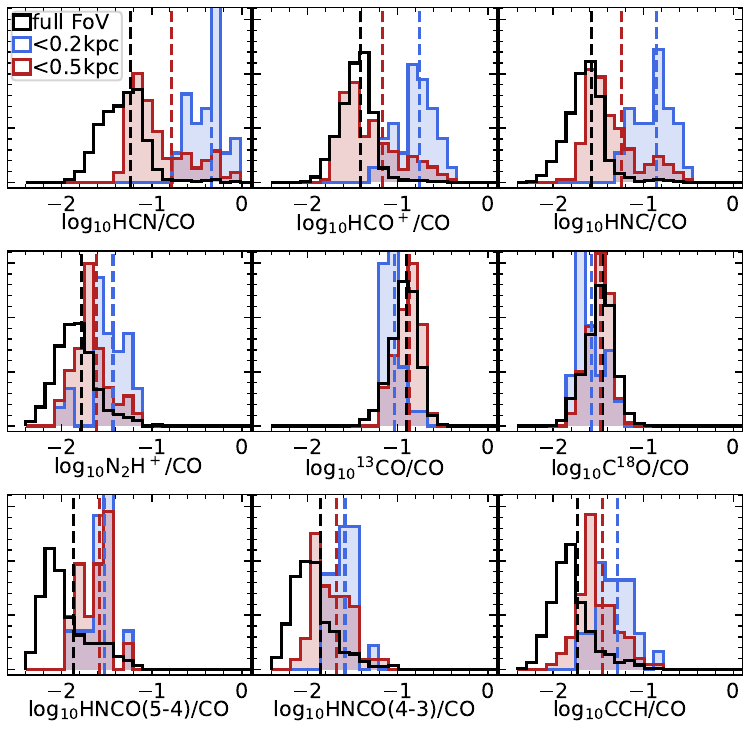}
    \caption{Histograms of line ratios with \CO for pixels where both the line emission and \CO emission are significantly detected  ($>3\sigma$) inside the full FoV (black) and the central 1\,kpc (blue) and central 0.4\,kpc (in diameter, red). We mark averages (log$_{10}$(mean(line/CO))) for the full FoV and central apertures (grey, blue and red dashed line, respectively). The histograms are normalized to have an integrated area of unity.}
    \label{fig:Histograms_LinevsCO}
\end{figure}

\CO is often used to study the bulk molecular gas distribution in galaxies \citep[i.e.,][]{helfer_bima_2003, bolatto_co--h2_2013,leroy_phangsalma_2021}, because it is relatively abundant and its rotational transitions easily excited,  producing bright millimeter-wavelength emission under typical ISM conditions. 
When studying other usually fainter molecular lines, it is thus often of interest to measure their intensity relative to \CO. 
In Figure~\ref{fig:Histograms_LinevsCO}, we show the logarithmic distribution of the emission in the SWAN moment-0 maps divided by the integrated intensity \CO emission  \citep[from ][see Section~\ref{sec:m0mapcreation}]{schinnerer_pdbi_2013}. 
The distribution of the integrated intensity ratios are shown separately for the full disk and the central 1\,kpc and central 0.4\,kpc. 
Here, full disk refers to the area inside the SWAN FoV, where we avoid the increased noise towards the edges of the mosaic (compare Figure~\ref{fig:mosaicpointings}). 
We only consider pixels where the respective line and CO are significantly detected ($>3\sigma$). 
Table~\ref{tab:LineratiosCO} provides the  average integrated intensity line ratios in the full disk, central 1\,kpc and disk excluding the center, as well as their scatter. 
To assess the effect of the 3-sigma masking on the total flux, we also quote the fraction of masked-to-unmasked flux.
Even within the environments, we note large variations in the \CO line ratios (Figure~\ref{fig:Histograms_LinevsCO}). We estimate how much noise is contributing to this scatter in Section~\ref{sec:Science:LinevsCOScatter}.

As before, the central 1\,kpc distribution is clearly offset from the full disk distribution for most lines except for the \CO line ratios with the CO isotopologues.
This offset is even stronger for the HCN-to-CO, \hcop-to-CO, HNC-to-CO and \cch-to-CO ratio in the central 0.4\,kpc (diameter).
This is caused by \CO emission being nearly absent in the very center of the galaxy, presumably due to photodissociation, mechanical evacuation, or radiative
transfer effects \citep{querejeta_agn_2016}.
The similar behavior of \tco-, and \ceto-to-\CO (1-0) ratios is consistent with results from SMA observations \citep[SMA-PAWS; ][]{den_Brok_2024SWAN_SMA}, which cover the (2-1) transitions of these CO isotopologues. 
The similar radial trends observed in the \tco-to-\CO (1-0) and (2-1) ratios suggest that these transitions are influenced by similar excitation mechanisms. 
Still, by integrating the SWAN and SMA-PAWS CO isotopologue line observations, the non-LTE modeling analysis in \citet{den_Brok_2024SWAN_SMA} indicates opacity variations at cloud scales within the disk. 
To assess possible effects of increased \CO opacity, we provide histograms of line ratios with \tco in Appendix~\ref{app:13coratios}. Qualitatively the same increases for the dense gas tracer to CO line ratios in the central enhancements can be seen.

We will discuss this in more detail in Section~\ref{sec:Discussion}.
Since both \ceto and \tco show a similar lack of emission in the center as \CO, their distributions of center and disk agree well. 

On average, HCN emission is $\sim17$ times fainter than \CO, but this factor varies strongly between $\sim 6$ in the galaxy center, and $\sim20$ in the disk. 
This emphasizes the importance of mapping emission not only in individual environments but across a larger set of environments such as in our SWAN FoV. 
The faintest lines detected in our data, \nnhp{}(1-0) and HNCO(4-3), are roughly a factor of $\sim70$ times fainter than \CO in the disk.

\begin{table} 
\begin{small}
\tabcolsep=0.11cm
\caption{Typical line ratios with \CO}\label{tab:LineratiosCO}
\centering
\begin{tabular}{l|lll|ll}
\hline\hline
\noalign{\smallskip}
    Line ratio &  full disk & center & disk & F$_\mathrm{line}$ & F$_\mathrm{CO}$ \\
    \noalign{\smallskip}
\hline
\noalign{\smallskip}
HCN(1-0)/CO & $0.059 \pm ^{0.027}_{0.009}$ & $0.166 \pm ^{0.101}_{0.015}$ & $0.049 \pm ^{0.018}_{0.015}$ & 0.95 & 0.83 \\
HNC(1-0)/CO & $0.027 \pm ^{0.010}_{0.003}$ & $0.058 \pm ^{0.029}_{0.001}$ & $0.022 \pm ^{0.006}_{0.005}$ & 0.87 & 0.66 \\
HCO$^+$(1-0)/CO & $0.039 \pm ^{0.012}_{0.006}$ & $0.067 \pm ^{0.039}_{0.006}$ & $0.037 \pm ^{0.01}_{0.008}$ & 0.93 & 0.78 \\
N$_2$H$^+$(1-0)/CO & $0.017 \pm ^{0.007}_{0.002}$ & $0.024 \pm ^{0.008}_{0.002}$ & $0.016 \pm ^{0.006}_{0.002}$ & 0.75 & 0.45 \\
HNCO(5-4)/CO & $0.014 \pm ^{0.007}_{0.001}$ & $0.026 \pm ^{0.010}_{0.007}$ & $0.011 \pm ^{0.004}_{-0.0}$ & 0.74 & 0.09 \\
HNCO(4-3)/CO & $0.014 \pm ^{0.006}_{0.001}$ & $0.021 \pm ^{0.008}_{0.006}$ & $0.013 \pm ^{0.005}_{0.010}$ & 0.68 & 0.25 \\
C$^{18}$O(1-0)/CO & $0.035 \pm ^{0.009}_{0.007}$ & $0.034 \pm ^{0.007}_{0.006}$ & $0.035 \pm ^{0.009}_{0.007}$ & 0.89 & 0.77 \\
$^{13}$CO(1-0)/CO & $0.125 \pm ^{0.028}_{0.024}$ & $0.137 \pm ^{0.028}_{0.028}$ & $0.124 \pm ^{0.028}_{0.024}$ & 0.98 & 0.96 \\
\cch{}(1-0)/CO & $0.019 \pm ^{0.007}_{0.001}$ & $0.035 \pm ^{0.012}_{0.006}$ & $0.016 \pm ^{0.005}_{0.001}$ & 0.55 & 0.32 \\
    \hline 
    \noalign{\smallskip}

\end{tabular}
\tablefoot{Average line ratios with \CO(1-0) inside the full FoV (limited to the area covered by the mosaics, see Figure~\ref{fig:mosaicpointings}, the central 1\,kpc (in diameter) and the disk excluding the central 1\,kpc. Average line ratios are determined from pixels in the moment-0 maps where both lines are detected significantly ($>3\sigma$). We provide the difference between the average and the 25$^\mathrm{th}$ and 75$^\mathrm{th}$ percentiles as uncertainty. 
To estimate the fraction of emission that originates from regions of significant detection, we provide the masked-to-unmasked flux ratio for both the specific emisison line (F$_\mathrm{Line}$) and CO (F$_\mathrm{CO}$).
This ratio is calculated by comparing the total flux in the full FoV within regions were the flux exceeds the $>3\sigma$ threshold (masked flux) to the total flux without applying any threshold (unmasked flux). 
}
\end{small}
\end{table}

\subsection{Effect of noise on measured line ratios with \CO}
\label{sec:Science:LinevsCOScatter}

In addition to the variations in \CO line ratios between center and disk, we observe a large scatter in the histograms presented in Figure~\ref{fig:Histograms_LinevsCO}, reflected by the large percentiles in Table~\ref{tab:LineratiosCOScatter}). 
Our goal is to determine whether the scatter in our dataset can be attributed solely to noise, or if it might be linked to physical properties.
The 3-sigma clipping threshold applied during the analysis  can introduce a bias in the measurement of line ratios and their scatter. 

To explore this, we built a simple toy model for which we assume that the line-to-CO ratio remains constant, and is represented by the median line ratio, which we provide in Table~\ref{tab:LineratiosCOScatter}. This median line ratio is calculated for pixels where both the line and \CO are signficantly detected (>3$\sigma$), similar to the mean values in Table~\ref{tab:LineratiosCO}).
In addition, we provide the median absolute deviation $s_\mathrm{measured}$, which we use as indication of the scatter. 
For each pixel we predict the expected line intensity based on the \CO intensity and the constant line ratio. 
We add a Gaussian noise distribution which is based on each pixels noise estimate on both the predicted line intensity and the \CO intensity.  
Next, we calculate the scatter of the line ratio of these two modified intensities.  
This procedure is repeated 100 times and the median scatter $s_\mathrm{calculated}$ is quoted in Table~\ref{tab:LineratiosCOScatter}. 
For all line ratios, the calculated scatter is several times smaller than the measured one. 
This implies that the measured scatter cannot be fully explained by noise and physical mechanisms might be contributing to the scatter. 

We estimate the physical contribution by assuming a linear dependency of the line ratio and a statistical and physical noise contribution. This leads to $s_\mathrm{physical} = \sqrt{ s_\mathrm{measured}^2 - s_\mathrm{calculated}^2}$. We list $s_\mathrm{physical}$ for all lines in Table~\ref{tab:LineratiosCOScatter}.
We find that noise uncertainties make a minor contribution to the scatter measured for all line ratios, meaning that the variations in line ratios that we measure at 125\,pc resolution are mostly driven by physical mechanisms.

The simple assumption of a constant line ratio is not correct for all lines. 
As shown by S23, \nnhp depends superlinearly on \CO (power of $\sim1.1$) and HCN depends sub-linearly on \CO (power of $\sim0.5$). 
Since most of these trends are driven by a small number of brighter pixels (i.e. surrounding the AGN), their contribution to the scatter is small. 
To first order, we consider this a reasonable assumption. We conclude that the variations observed are the effect of physical properties changing and not due to variations associated with the noise of the images.

\begin{table}
\begin{tabular}{ccccc}

\end{tabular}
\end{table}
\begin{table} 
\begin{small}
\tabcolsep=0.11cm
\caption{Typical scatter of median \CO line ratios}\label{tab:LineratiosCOScatter}
\centering
\begin{tabular}{l|llll}
\hline\hline
\noalign{\smallskip}
    Line ratio & full disk median & $s_\mathrm{measured}$ & $s_\mathrm{calculated}$ & $s_\mathrm{physical}$ \\
    \noalign{\smallskip}
\hline
\noalign{\smallskip}
HCN(1-0)/CO & 0.049 & 0.026 & 0.005 & 0.025 \\
HNC(1-0)/CO & 0.023 & 0.009 & 0.003 & 0.009 \\
HCO$^+$(1-0)/CO & 0.036 & 0.013 & 0.004 & 0.013 \\
N$_2$H$^+$(1-0)/CO & 0.014 & 0.006 & 0.003 & 0.006 \\
HNCO(5-4)/CO & 0.009 & 0.004 & 0.002 & 0.003 \\
HNCO(4-3)/CO & 0.011 & 0.004 & 0.002 & 0.004 \\
C$^{18}$O(1-0)/CO & 0.033 & 0.012 & 0.005 & 0.011 \\
$^{13}$CO(1-0)/CO & 0.121 & 0.039 & 0.011 & 0.038 \\
\cch{}(1-0)/CO & 0.015 & 0.006 & 0.003 & 0.005 \\
    \hline 
    \noalign{\smallskip}

\end{tabular}
\tablefoot{Median line ratios with \CO(1-0) inside the full FoV calculated from pixels where both lines are detected significantly ($>3\sigma$, see Table~\ref{tab:LineratiosCO}). 
We list their median absolute deviation to estimate the scatter ($s_\mathrm{measured}$) and calculate the expected scatter based on the uncertainty maps and the median line ratio ($s_\mathrm{calculated}$). 
The calculation of $s_\mathrm{calculated}$ is described in the text. 
Under the assumption that the measured line ratio consists of a constant line ratio with statistical noise and a physical contribution, we can estimate the scatter due to physical effects as $s_\mathrm{physical} = \sqrt{s_\mathrm{measured}^2 - s_\mathrm{calculated}^2}$.
}
\end{small}
\end{table}

\subsection{Comparison of SWAN data with other surveys}
\label{sec:Science:Literature}

Here we compare the SWAN results with previous observations to test for (a) consistency with high-resolution ($\sim3\arcsec$) observations of HCN(1-0) in three individual pointings in M51 \citep{querejeta_dense_2019}, as well as (b) put the data into context with lower-resolution ($\sim 30\arcsec$), but larger-FoV observations of HCN, \hcop and HNC from the EMPIRE survey \citep{jimenez-donaire_empire_2019} and \tco, \ceto and \nnhp from the CLAWS survey \citep{den_brok_co_2022}. 
While the PyStructure table is well suited for the line-by-line comparison done before, the GILDAS moment-maps and cubes are best suited for such a comparison with literature works (compare Section~\ref{sec:m0mapcreation}).

\subsubsection{Comparison with high-resolution HCN observations}

\begin{figure}
    \centering
    \includegraphics[width = 0.242\textwidth]{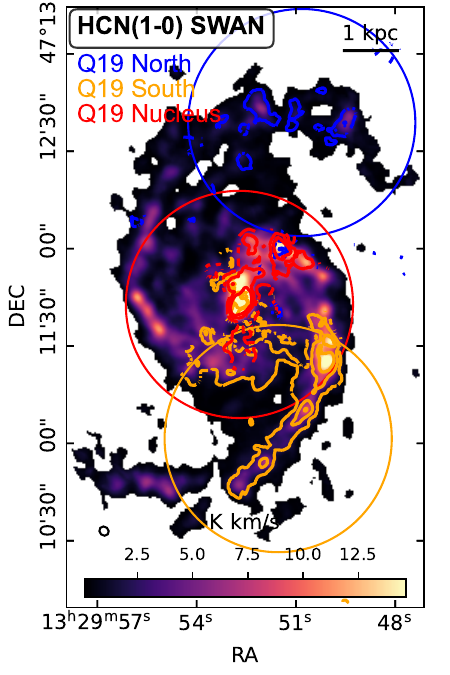}
    \includegraphics[width = 0.242 \textwidth]{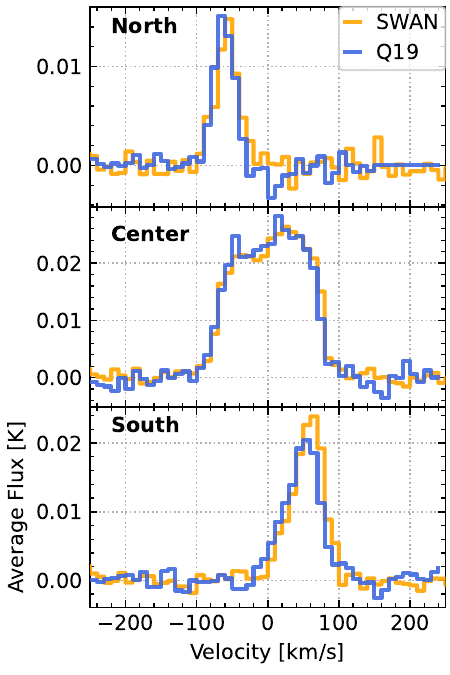}
    \caption{Comparison with HCN observations from \citet{querejeta_dense_2019} (Q19). Left panel: SWAN HCN moment-0 map with contours ($\sim5,50,200\sigma$ with $\sigma$ the average noise) of Q19 HCN maps. Both maps are at 3\arcsec{} resolution, spatially and spectrally regridded to the same grid, and the maps are created via the same GILDAS island method (using just the HCN line). Circles depict radii of 35\arcsec{} centered according to Q19. Right panels: Average spectra of SWAN 3\arcsec{} data at 10\,km\,s$^{-1}$ resolution as well as the matched Q19 data. Spectra are the average flux inside of the circular area shown in the left panel.}
    \label{fig:HCNMiguel}
\end{figure}

HCN has previously been observed in M51 at a similar resolution (3\arcsec) with the IRAM 30m and PdBI interferometer by \citet{querejeta_dense_2019} (Q19) in three pointings in the disk. 
This survey mapped the HCN(1-0) flux inside 35\arcsec-sized pointings centered on M51's center (RA = 13:29:52.708, Dec = +47:11:42.81 (J2000)), the northern spiral arm (RA = 13:29:50.824,
Dec = +47:12:38.83(J2000)) and the southern spiral arm (RA = 13:29:51.537, Dec = +47:11:01.48(J2000)). 
We convolve the data to a spatial (3.04\arcsec) and spectral (10\,km\,s$^{-1}$) resolution matched to our SWAN data. 
The data set from Q19 is integrated via the same GILDAS island method that we used for SWAN (Section~\ref{sec:m0mapcreation}).
Figure~\ref{fig:HCNMiguel} indicates the HCN observations by Q19 as contours overlaid on our SWAN HCN(1-0) map, as well as average spectra from both datasets corresponding to the 35\arcsec{} apertures. 

We find good agreement between both the spatial and the spectral distribution of the HCN emission in the Q19 and SWAN data sets. 
The total emission integrated over the matched spectra in the northern region of Q19 represents $\sim119\%$ of the  total SWAN emission integrated over the spectra inside the same region. 
This is likely due to the northern region of Q19 slightly exceeding the SWAN FoV and thus capturing a slightly larger area. 
For the center, this fraction is $\sim96\%$ and for the southern pointing $\sim91\%$. 

\subsubsection{Comparison of HCN, \hcop, and HNC emission from EMPIRE and CLAWS}

\begin{figure}
    \centering
    \includegraphics[width = 0.5\textwidth]{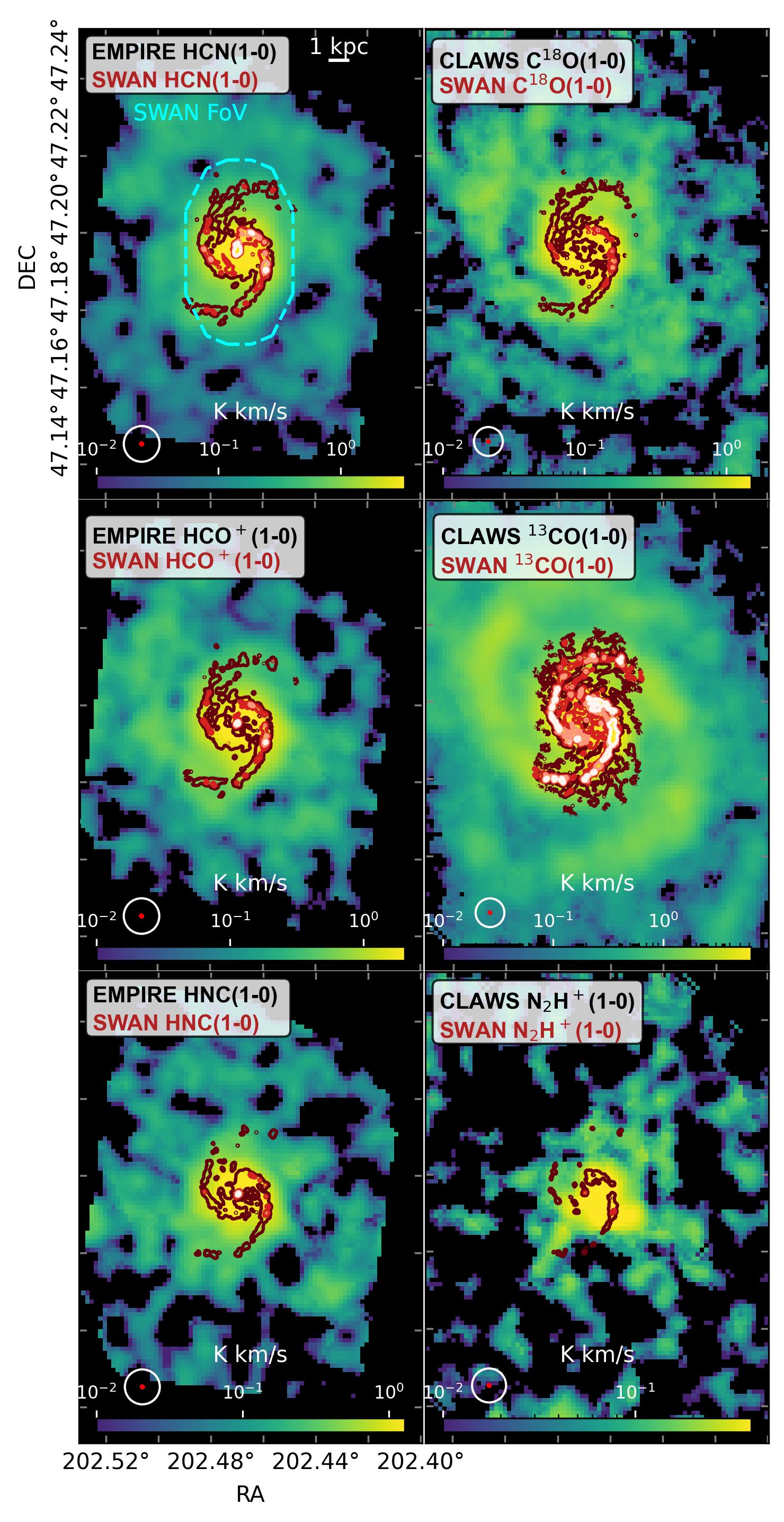}
    \caption{Integrated emission maps (moment-0) from EMPIRE at $33\arcsec$ for HCN, \hcop, HNC (1-0) (left columns) as well as from CLAWS at  15-30\arcsec{}resolution for \ceto, \tco and \nnhp (1-0) (right columns) with SWAN 1,5,10 and 15 K km\,s$^{-1}$ contours at native resolution ($\sim 3 \arcsec$) on top. Beam sizes of EMPIRE/CLAWS and SWAN are shown in the bottom left corner of each panel (white and red circles, respectively)}
    \label{fig:EMPIRE_gallery}
\end{figure}

Observations of HCN, \hcop and HNC at both lower and higher resolution have been proven crucial to study the conditions of molecular gas \citep{helfer_dense_1993,aalto_variation_1997,meierturner2005, kohno_prevalence_2005,  bigiel_empire_2016, jimenez-donaire_empire_2019, den_brok_co_2022, imanishi_dense_2023, neumann_almond_2023, nakajima_molecular_2023}. 
To showcase the difference in resolution, Figure~\ref{fig:EMPIRE_gallery} depicts $\sim$kpc observations from EMPIRE for HCN, \hcop, and HNC(1-0) emission, as well as from CLAWS for \ceto, \tco, and \nnhp{}(1-0), and SWAN contours on top. 
While the EMPIRE/CLAWS maps cover the outer parts of M51 better due to their larger FoV, their coarse resolution misses several structures that we can resolve in SWAN. 
SWAN shows a clear difference in molecular line emission between the spiral arms and interarm regions. 
Additionally, in contrast to EMPIRE/CLAWS, in SWAN we can differentiate between the molecular ring and the AGN-impacted center, which are regions with very different physical conditions.

\section{Discussion}
\label{sec:Discussion}

The SWAN survey provides a view of CO isotopologues, dense gas tracing molecular emission lines and PDR and shock-tracing lines at the sensitivity and spatial resolution (125\,pc) required to bridge extragalactic and Galactic studies. 
The $\rm 5\times7\,kpc^2$ FoV of SWAN covers the central region, which hosts an AGN, a nuclear bar, and a molecular ring, as well as spiral arms and the interarm region. 
Although all lines are prone to different excitation conditions, our high-resolution maps show general similarities between their distributions across the FoV (Figures~\ref{fig:Gallery_overview}, \ref{fig:Gallery} and \ref{fig:Gallery_GILDAS}).  
All lines are detected along the northern to south-western side of the molecular ring, with the brighter lines extending well along the spiral arms.  
This work provides a first analysis of this dataset and we report stark differences between the emission of these lines and their line ratios with \CO between the central 1\,kpc and disk.

\begin{figure}
    \centering
    \includegraphics[width = 0.45\textwidth]{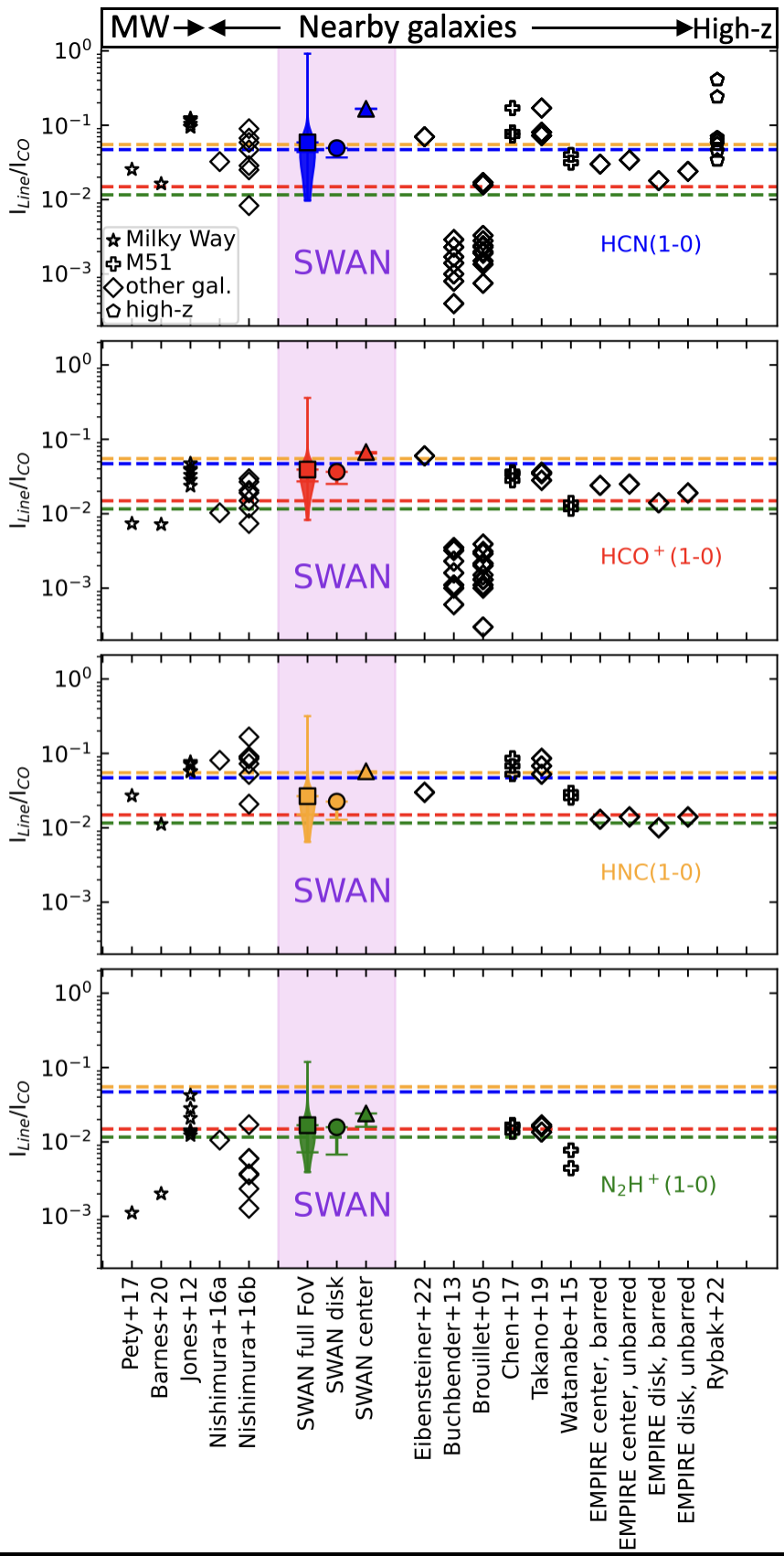}
    \caption{Literature comparison of integrated intensities for the J=1-0 transition of (from top to bottom:) HCN, HNC, \hcop, and \nnhp emission compared to \CO. We show the SWAN integrated emission (light-shaded area) for the full FoV (squares), the central 1\,kpc (triangle), and the remaining disk (circle). These values are obtained by integrating emission from pixels where both the line emission and \CO emission is significantly detected ($>3\sigma$). The distribution of pixels in the full FoV is added as violins. 
    The error bars correspond to the difference between this calculation (both \CO and the line are significantly detected) and when instead using pixels where \CO is significantly detected. 
    Average literature values (dashed horizontal lines, calculated in linear scale)  are based on values from Milky Way \citep{pety_anatomy_2017, barnes_lego_2020, jones_spectral_2012} and extragalactic sources: M51 studies at lower resolution \citep{watanabe_spectral_2014} and in the outer spiral arm \citep{chen_dense_2017}; different galaxy averages from EMPIRE \citep{jimenez-donaire_empire_2019}; $\sim100$pc studies in M33 \citep{buchbender_dense_2013}, M31 \citep{brouillet_hcn_2005} and NGC6946 \citep{eibensteiner_23_2022}; $\sim10\,$ pc observation in the LMC \citep{nishimura_spectral_2016-1} and $80\,$pc in the dwarf galaxy IC10 \citep{nishimura_spectral_2016}; $\sim15-19\arcsec$ in three nearby galaxies \citep{takano_molecular_2019} and upper limits from $z\sim3$ galaxies \citep{rybak_prussic_2022}. The values are basically sorted by their physical resolution from a few pc in the Milky Way (left) to several kpc in external galaxies (right).}
    \label{fig:LiteratureDense}
\end{figure}

\begin{figure}
    \centering
    \includegraphics[width = 0.5\textwidth]{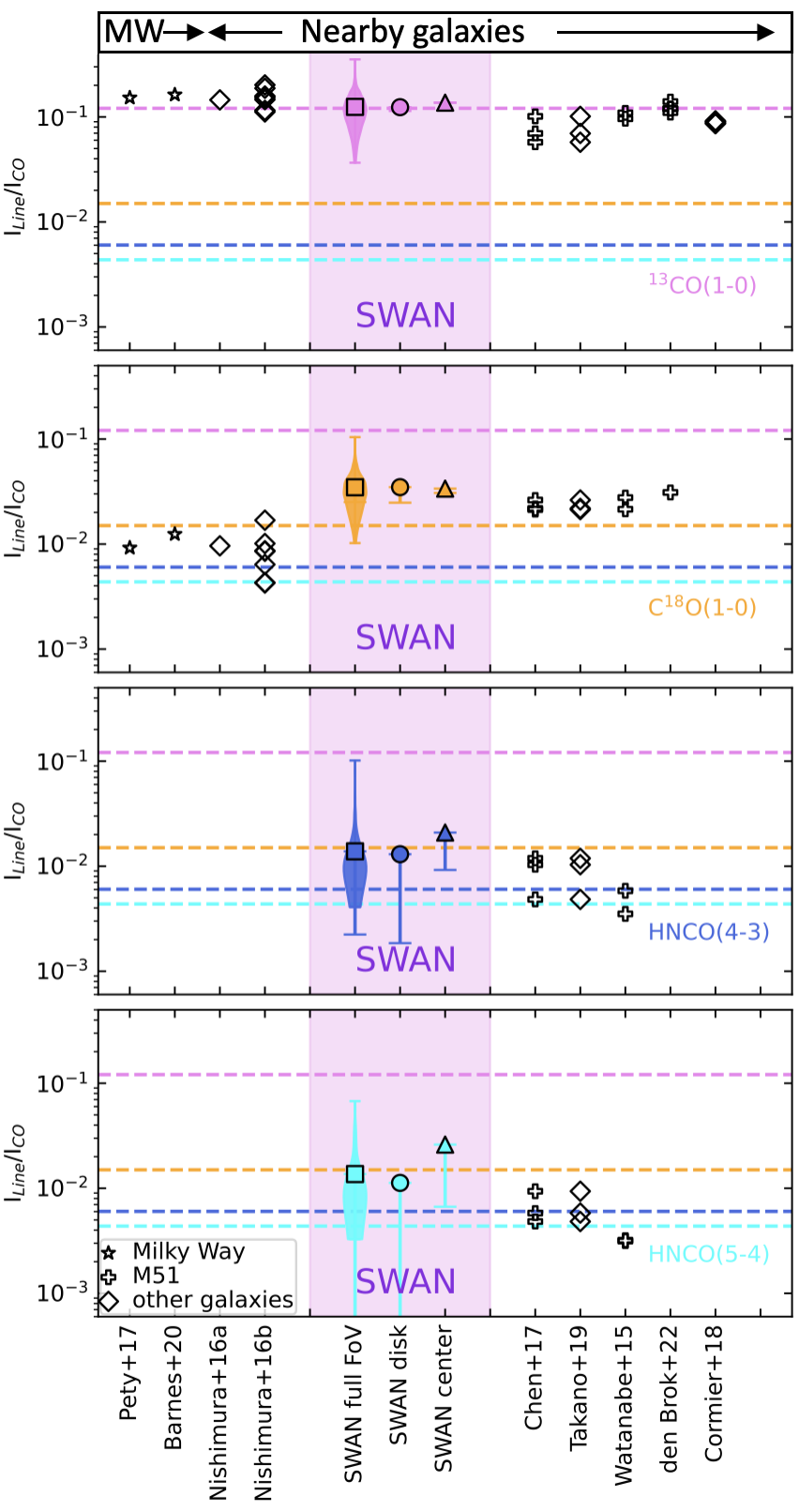}
    \caption{Same as Figure~\ref{fig:LiteratureDense}, but for the J=1-0 transition of the CO isotopologues \ceto, and \tco, as well as the HNCO(5-4) and HNCO(4-3) lines. The EMPIRE isotopologues are measured by \citet{cormier_full-disc_2018} instead of the survey paper. Further, we add CLAWS measurements of M51 \citep{den_brok_co_2022}. The measurements are sorted by physical resolution (increasing to $\sim$ kpc-scales at the right side of the plot).}
    \label{fig:LiteratureIsotopo}
\end{figure}

CO emission is commonly used to trace the bulk molecular gas distribution. 
Hence, line ratios with \CO are used to gauge the abundance of molecules or estimate the average gas density \citep{leroy_millimeter-wave_2017, jimenez-donaire_empire_2019}. 
We compare our average CO line ratios from center, disk and full FoV to literature values from Milky Way to high-redshift studies in Figure~\ref{fig:LiteratureDense} and Figure~\ref{fig:LiteratureIsotopo}.
The error bars indicate the change in the average line ratio when applying different masking. When instead of selecting pixels where both the line and \CO are significantly detected (Table~\ref{tab:LineratiosCO}), we only require \CO to be detected, the average line ratio with \CO decreases for most lines. 
The distribution of significantly detected pixels is shown for the full FoV as violins for visual comparison. 
We discuss the variations of line intensities and CO line ratios within M51 and in comparison to other galaxies below.

\paragraph{CO isotopologues \tco and \ceto:}

The general deficit of the otherwise abundant \CO, \ceto and \tco emission in the center (Figure~\ref{fig:Gallery_overview}, ~\ref{fig:Histograms_LinevsCO}) might suggest the chemical or mechanical destruction or excitation of these molecules via the jet and associated mechanisms. 
The former is in agreement with findings by \citet{saito_agn-driven_2022} in the outflow of NGC~1068, where \CO isotopologues are faint and potentially destroyed by dissociating photons and electrons \citep[see also][]{cecil_spatial_2002}.

In general, all SWAN average line ratios between \CO and \tco or \ceto agree well with estimates for the Milky Way and nearby galaxies from the literature
(Figure~\ref{fig:LiteratureIsotopo}). 
Since \tco is significantly brighter than the other molecules (i.e., \nnhp) studied here, masking methods affect less the obtained ratio (compare F$_\mathrm{line}$, Table~\ref{tab:LineratiosCO}). 
Still, a slight decrease in line ratio with increased physical resolution might be inferred from Figure~\ref{fig:LiteratureIsotopo}, but only a few data points are available. 
Further, we do not see a large difference in line ratios between the center and disk of M51 that is evident for the dense gas tracers. 
This is in contrast to findings at 100\,pc scales in the active galaxy NGC~3627 \citep{beslic_dense_2021} where the \tco-to-\CO(2-1) ratio is decreased in the center with respect to outer regions such as bar ends. Since NGC~3627 hosts a larger bar than M51 does, the central dynamics might play a critical role. 

The \ceto-to-\CO ratio increases with increasing physical resolution (Figure~\ref{fig:LiteratureIsotopo}, studies roughly sorted by physical resolution) which indicates that masking might affect the measurements. 
Interestingly, the SWAN measurements are generally higher than measurements from individual regions in both the Milky Way and other galaxies and are comparable to the kpc measurements from CLAWS \citep{den_brok_co_2022} which cover a larger FoV including the entire molecular gas dominated disk of M51.

\paragraph{Shock tracer HNCO:}
The higher rotational transitions of HNCO in the 3\,mm range have been suggested as tracer of low-velocity shocks \citep[e.g.,][]{Martin_2008ApJ, Kelly_2017A&A}.
HNCO emission thus might suggest the presence of shocks in M51's center consistent with the location of a low-inclination radio jet and a dense-gas outflow containing both \CO and HCN emission \citep{querejeta_dense_2019}. 
This is in agreement with findings by \citet{Martin_2015} who report enhanced HNCO emission in the cicumnuclear disk surrounding the AGN in NGC~1097 at $\sim100\,$pc resolution, and observations of HNCO arising from two lobes at both sides of the AGN of Seyfert galaxy NGC~1068 \citep{Takano_2014}.

The average HNCO(4--3)-to-\CO line ratios of M51's center and disk are comparable to the high-resolution study in the outer arm of M51 \citep{chen_dense_2017}, as well as the lower-resolution study in M51 \citep{watanabe_spectral_2014} and other galaxy measurements \citep[NGC~253, NGC~1068, and IC~342;][]{takano_molecular_2019}.
As HNCO(5-4) emission is only detected from very few sight lines, the errorbar on its average line ratio with \CO is very large  (Figure~\ref{fig:LiteratureIsotopo}).
Masking pixels without significant detections elevates the HNCO(5--4) average significantly (compare the error bar in Figure~\ref{fig:LiteratureIsotopo} which showcases the average when including non-detections).
With only 9\% of the total \CO flux found in the area where HNCO(5--4) is significantly detected (Table~\ref{tab:LineratiosCO}), we caution the use of this masked full FoV HNCO(5--4)-to-\CO ratio.

\paragraph{Dense gas tracing lines - HCN, HNC, \hcop, \nnhp:}
While some molecules such as \ceto and \tco (and \CO) are very faint in the center, other lines, such as HCN(1-0), are very bright (Figure~\ref{fig:Lineemissioncomparison}, Table~\ref{tab:offsetb}) and enhanced compared to \CO  emission (Figure~\ref{fig:Histograms_LinevsCO}, Table~\ref{tab:LineratiosCO}) or \tco emission ( Figure~\ref{fig:Histograms_Linevs13CO}).  This enhancement is increased when considering even smaller central apertures, suggesting that the outflow and/or AGN could be the potential cause (Figure~\ref{fig:Histograms_LinevsCO}).

M51 has long been known to exhibit increased HCN emission in the very center both from kpc-scale studies \citep{jimenez-donaire_empire_2019} and that at $\rm 30$ and $100\,$pc resolution \citep{helfer_dense_1993, matsushita_resolving_2015, querejeta_dense_2019}. 
Infrared pumping, weak HCN masing, an increased HCN abundance or electron excitation in the X-ray dominated region (XDR) of the AGN are some possibilities suggested throughout the literature \citep[e.g.,][]{blanc_spatially_2009,matsushita_resolving_2015,querejeta_agn_2016, goldsmith_electron_2017, stuber_surveying_2023}. 
HCN might potentially partake in the outflow driven by M51's AGN and radio jet. 
Similarly, bright HCN emission is co-located at the location of the AGN-driven outflow in the center of the galaxy merger NGC~3256 \citep{michiyama_alma_2018, harada_alma_2018}, in Mrk~231 \citep{aalto_detection_2012}, starburst galaxy NGC~251 \citep{beslic_dense_2021} and NGC~1068 \citep{saito_agn-driven_2022} 
and in the center of NGC~4321 \citep{neumann_ngc4321_2024}.
While both NGC~1068 and M51 host a small weak radio jet, there are differences: no \nnhp emission is detected from the outflow in NGC~1068 \citep{saito_agn-driven_2022}, while bright \nnhp emission is evident in M51's center and outflow, with increasing average \nnhp-to-CO ratios for smaller central apertures (Figure~\ref{fig:Histograms_LinevsCO}). 

Shocks might also be an effective way to destroy or excite \CO molecules to higher states. 
A potentially young or weak jet might not yet have been able to mechanically remove large quantities of molecular gas. 
We might be viewing an early stage of molecular gas destruction, where lower density molecular gas such as \CO, which usually covers a larger volume than i.e., HCN molecules, is destroyed in large quantities, while HCN, which is brighter in denser regions (smaller volume) is not yet affected, or even shock-enhanced. (i.e., weak maser or abundance increase). 
However, dynamical age estimates of the jet or central objects in M51 range from $\sim10^4-10^5$yr \citep{ford_bubbles_1985, matsushita_jet-disturbed_2007} up to a few Myrs \citep{rampadarath_jets_2018}.
The destruction of \CO might allow molecules such as \nnhp to form more abundantly as they would otherwise rapidly react with \CO \citep{bergin_cold_2007}. 
Dissociation of other molecules by the radio jet might provide a large quantity of free electrons in the center of M51, which then leads to an efficient dissociative recombination of \nnhp. 
The exact mechanisms driving the line emission in M51's center will be discussed in more detail in Thorp et al. (in prep.) and Usero et al. (in prep.).

Similarly to M51, \citet{meierturner2005} find bright \nnhp emission in the center of IC342 at 5$\arcsec$, which they explain by an increased cosmic ray ionization rate (CRIR) or an enhanced N$_2$ abundance. 
An increased \nnhp abundance could then promote the formation of \hcop which can form from \nnhp molecules \citep{harada_molecular-cloud-scale_2019}. 
This is consistent with the fact that \hcop emission in the center follows a significantly different relation with HCN emission and is more tightly related to \nnhp than HCN (compare Appendix~\ref{app:Offsets}). This is in agreement with findings by \citet{butterworth_molecular_2024} that \hcop and its isotopologues have larger column densities than HCN in the starbursting center of NGC~253 from the ALMA Comprehensive High-resolution Extragalactic Molecular Inventory (ALCHEMI) survey \citep{martin_alchemi_2021}. 
Given that neither IC342, nor NGC~253 host an AGN, and M51 does not host a starburst in its center, the similarities between these galaxy centers are puzzling.

Emission from HCN and its isomer HNC are tightly correlated and show only a weak difference in their relation between the center and disk, suggesting that the chemical conditions able to convert one molecule into the other are not changing between the two environments. 
This is in agreement with findings by \citet{meierturner2005} in IC342. 
A detailed study of the cloud-scale variations of the HCN-to-HNC ratio will be presented in Stuber et al. (in prep.).

Figure~\ref{fig:LiteratureDense} shows a large spread between the average ratios of the dense gas tracers to \CO from the literature and SWAN, especially for HCN, HNC and \hcop. 
Milky Way studies at high resolution report lower values compared to SWAN, except for the CMZ \citep{jones_spectral_2012} which is consistent with our central averages being increased for all lines. 
The variation between multiple lines per literature study (i.e., average HCN/CO ratio compared to the \hcop/CO line ratio for a single study) is typically smaller than the variation of one line ratio across all studies (e.g., HCN/CO varies strongly from study to study). 
This might indicate that different masking techniques drastically change the resulting ratios. 
While some studies apply cloud-finding algorithms to isolate individual clouds, other studies quote full FoV averages. 
Studies at kpc-resolution such as \citet{watanabe_spectral_2014} in the center and south-western molecular ring in M51 and the EMPIRE survey \citep{jimenez-donaire_empire_2019} report in general lower line ratios, and so do high-resolution Milky Way studies.  
Generally, we find no clear trend with resolution (Figure~\ref{fig:LiteratureDense}, studies are basically sorted by resolution). 

As can be seen from Figure~\ref{fig:Histograms_LinevsCO}, the pixel-by-pixel distribution of the line ratios with \CO generally spans over at least one or even two orders of magnitude for most lines. 
In Section~\ref{sec:Science:LinevsCOScatter} we estimated how much of the scatter can be attributed to noise and found that there is significant scatter in all line ratios with \CO that can not be explained by noise only. 
Several studies report dependencies of ratios between dense gas tracers and \CO on dynamical equilibrium pressure or stellar surface density \citep{usero_variations_2015,querejeta_dense_2019, neumann_almond_2023}. 
A strong radial dependency is found in EMPIRE \citep{jimenez-donaire_empire_2019}, and their line ratios with \CO are generally enhanced in galaxy centers compared to disks (Figure~\ref{fig:LiteratureDense}). 
Still, the difference between their centers and disks is smaller than our SWAN variations between center and disk, despite EMPIRE covering a larger FoV. 
HCN, \hcop, and HNC emission in the center of NGC~6946 \citep{eibensteiner_23_2022} sit between our disk and center measurements, despite NGC~6946 not hosting an AGN. 
Elevated central line emission is a feature common to galaxies with and without an AGN \citep{usero_variations_2015, bigiel_empire_2016, gallagher_dense_2018, jimenez-donaire_empire_2019, heyer_dense_2022, neumann_ngc4321_2024} and attributed to the gas-rich high surface density common to galaxy centers. Studies at a similar resolution to ours in M\,33 and M\,31 both find significantly lower HCN and \hcop, which they consider a result of the sub-solar metallicity in both M\,33 and M\,31 \citep{buchbender_dense_2013}. 
The variation of dense gas tracers \nnhp, \hcop, HCN, and HNC with various physical quantities on cloud-scales is investigated in detail by Stuber et al. (in prep).

Overall, we find significant differences that can arise between the center (where an AGN, outflow, and a nuclear bar are present) and the disk of M51, whereas the variations in the disk are more subtle, but significant (Section~\ref{sec:Science:LinevsCOScatter}). 
The exact mechanisms affecting the molecular gas in the center and in particular the outflow will be studied in more detail in forthcoming papers (A. Usero et al. in prep., M. Thorp et al. in prep.), we emphasize that all lines except \CO and its isotopologues show enhanced emission in the center. 
Further, we find large variations in literature \CO line ratios across various resolutions and targets. 
The least variations are seen for \tco-to-\CO, which is consistent across the literature. 
Both the selected environment (center compared to disk) as well as the masking methods likely influence the line ratios.

\section{Summary}
\label{sec:Summary}

We present the first-of-its-kind high-resolution ($\lesssim 125\,$pc), high-sensitivity map of 3mm lines covering an area of $\sim5\times7\,$kpc$^2$ in the inner disk of the Whirlpool galaxy. 
We detect emission from CO isotopologues \tco and \ceto{}(1-0), dense gas tracing lines HCN, HNC, \hcop and for the first time across such a large FoV \nnhp{}(1-0). 
In addition, we detect HNCO(5-4), (4-3) and hyperfine transitions of \cch{}(1-0) in the center and molecular ring of M51. 
Comparing the emission of those lines to each other and to \CO{}(1-0) emission from PAWS \citep{schinnerer_pdbi_2013} at matched resolution, we find the following:

\begin{enumerate}
    \item  The high-resolution maps show general structural similarities across all molecular lines: Bright emission of all molecular lines is detected along the western side of the molecular ring. 
    Emission of the brighter lines is well detected in the southern and northern spiral arm, and all lines except the CO isotopologues are bright in M51's center. 

    \item We calculate typical ratios of line brightness with \CO brightness inside the full FoV for pixels where each line is significantly detected. We find the highest value for the \tco-to-\CO ratio ($0.125$), followed by HCN-to-\CO($0.059$), \hcop-to-\CO($0.039$), \ceto-to-\CO($0.03$), HNC{}-to-\CO($0.027$), \cch{}-to-\CO ($0.019$), \nnhp{}-to-\CO($0.017$) and HNCO{}-to-\CO(5-4: $0.014$, 4-3: $0.014$).  

    \item Emission of HCN is significantly enhanced in the central 1\,kpc compared to all other detected lines, and emission of \CO isotopologues is significantly reduced in the center compared to non-isotopologues. The only line combinations that do not exhibit a significant offset relation to each other between center and disk are the isotopologues \tco and \ceto, molecular ions \nnhp and \hcop, and combinations including the faint HNCO and \cch lines.  
    \CO line ratios are increased in the central 1\,kpc compared to the remaining disk for all lines except \ceto. The largest difference can be seen for the HCN-to-\CO line ratio, which is more than a factor of three times larger in the central 1\,kpc compared to the disk. 
    HNCO emission might suggest the presence of shocks in the galaxy center, linked to a low-inclined radio jet and a dense-gas outflow. 
    This points to complex conditions, possibly involving increased HCN abundance, infrared pumping, weak HCN masing, or electron excitation in the AGN's XDR.

    \item Line ratios with CO qualitatively compare well to lower-resolution literature studies in other galaxies, i.e., find a similar increase in CO line ratios for HCN, HNC, \hcop, and \nnhp in the centers of other galaxies. 
    Still, the overall spread in CO line ratios across Galactic and extragalactic sources is significantly larger than both the differences between center and disk in M51, and the differences between different lines (such as HCN, HNC, \hcop and \nnhp). 
    We find that the scatter of SWAN \CO line ratios can not be explained solely by noise and is likely attributed to local physical mechanisms.

\end{enumerate}

``Surveying the Whirlpool galaxy at Arcseconds with NOEMA'' (SWAN) allows us to study the molecular gas properties at cloud scale resolution across multiple environments in the iconic Whirlpool galaxy. 
The data is publicly available at the IRAM DMS\footnote{\url{https://oms.iram.fr/oms/?dms=viewobject/_=-O-404271&pageId=3}}. 
Dedicated studies focusing on the relationship of \CO isotopologues \citep{denBrok2025, Galic_submitted}, the dense-gas tracing lines (Stuber et al. in prep.), the outflow in the center of M51 (Usero et al. in prep., Thorp et al. in prep.) will showcase the utility of this rich data set for investigating physical conditions in the interstellar medium. 


\begin{acknowledgements}
This work has been carried out as part of the PHANGS collaboration and made use of data from the IRAM large program 'Surveying the Whirlpool galaxy at Arcseconds with NOEMA' (SWAN). 
We thank the referee for their valuable and constructive feedback. 
SKS acknowledges financial support from the German Research Foundation (DFG) via Sino-German research grant SCHI 536/11-1. 
JdB acknowledges support from the Smithsonian Institution as a Submillimeter Array (SMA) Fellow.
CE acknoledges the support of the Jansky Fellow of the National Radio Astronomy Observatory.
DL acknowledges the support from the Strategic Priority Research Program of the Chinese Academy of Sciences, grant No. XDB0800401.
HAP acknowledges support from the National Science and Technology Council of Taiwan under grant 110-2112-M-032-020-MY3.
AU acknowledges support from the Spanish grant PID2022-138560NB-I00, funded by MCIN/AEI/10.13039/501100011033/FEDER, EU.
\end{acknowledgements}

\bibliographystyle{aa} 
\bibliography{M51paper2.bib}

\begin{appendix}

\onecolumn

\section{Error beam contribution}
\label{app:errorbeam}

\begin{figure}
    \centering
    \includegraphics[width = 1\textwidth]{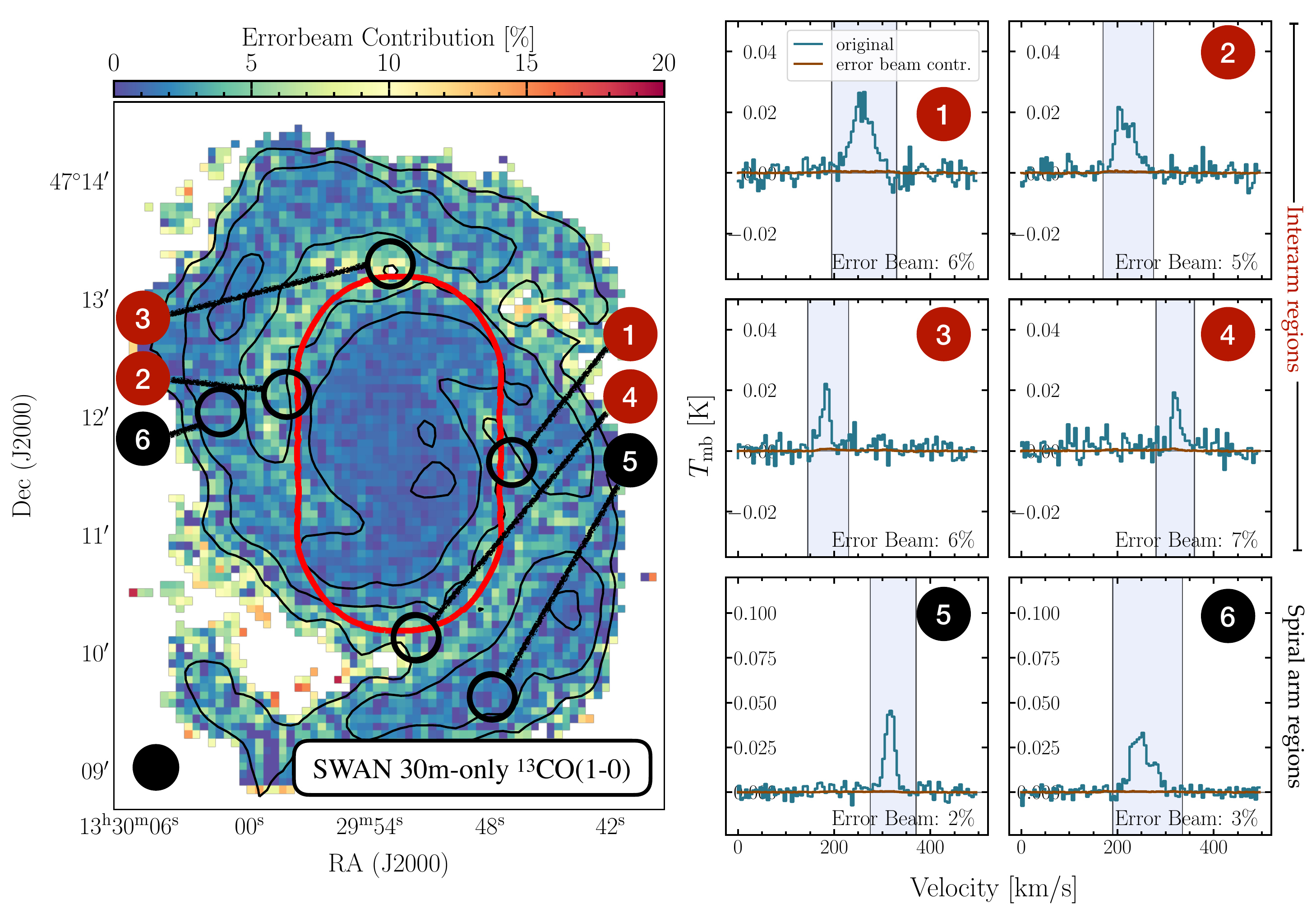}
    \caption{Quantifying errorbeam contributions to the observed $^{13}$CO(1--0) emission. (Left) The map presents the percentage errorbeam contribution to individual sightlines for the 30m only data. We only consider sightlines where the integrated $^{13}$CO(1--0) brightness temperature is detected at $\gtrsim3\sigma$. Contours indicate the SNR at 5,10,20, and 50. The red contour illustrates the SWAN field of view. Overall, the error beam contribution is marginal, with elevated values up to 10\% within the interarm region (and ${<}5\%$ within the SWAN field-of-view). (Right) We extract $^{13}$CO(1--0) spectra in six individual apertures. Apertures 1--4 are in the interarm region and apertures 5--6 in the spiral arms of M51. Each panel shows the full spectrum which includes the contribution from the error beam in blue. The brown spectrum represents the error beam contribution and is calculated by subtracting the error beam free spectrum from the observed spectrum. The percentage contribution is computed over the spectral range where we detect emission (indicated by the blue-shaded region) and is listed in each panel. }
    \label{fig:errorbeam}
\end{figure}

Our test show that the 30m error beams only make a small contribution to the flux filtered out by the interferometer.
We can describe the response of the 30m telescope to a point source by a set of 2D Gaussians, that represent the main beam and a set of error beams. As a result, depending on the morphology of the science target, the measured signal might be boosted by emission from beyond the region the telescope is pointing at as the wider side lobes pick up signals from other parts of the sky. For instance, in the case of M51, \citet{den_brok_co_2022} demonstrated that up to 20\% of the $^{12}$CO(2--1) emission in the interarm regions (which sit between two brighter spiral arms) can be accounted for by contributions from the error beams. 
\citet{Kramer2013} provide an approximation of the 30m telescope beam pattern at different frequencies, which we use as a good first-order approximation.

Given a model for the telescope's beam pattern, several deconvolution schemes exist that provide an estimate for the error beam free signal \citep[for example][]{Westerhout1973, Bensch1997, Lundgren2004, pety_plateau_2013, Leroy2015}. To quantify the relevance in the case of the set of lines observed by SWAN, we perform the procedure described in \citet{den_brok_co_2022} for $^{13}$CO(1--0), where we expect the effect to be the largest (as the main beam efficiency decreases with increasing frequency).  

\subsection{Mathematical framework}
We provide a brief overview of the mathematical framework and the method we use to deconvolve the 30m data to estimate the error beam contribution. For details on the calculations, we refer the reader to \citet{den_brok_co_2022}. The key parameters are the observed main beam temperature, $T_{\rm mb}$ (which includes also contributions from the error beam), and the error beam free main beam temperature, $\hat T_{\rm mb}$. They are related via a convolution kernel $K$ as follows:
\begin{equation}
    T_{\rm mb} = \left(\delta^{2\rm D }+ K\right)\otimes \hat T_{\rm mb},
\end{equation}
where ``$\otimes$" represents the 2D convolution operation and $\delta^{2\rm D }$ the Dirac 2D distribution. The kernel $K$ contains the sum of all error beam contributions after deconvolution with the main beam. 

The particular deconvolution, to obtain the error beam free signal, can be expressed using the Fourier Transform operation, $\mathcal{F}$:
\begin{equation}
    \hat T_{\rm mb} = \mathcal{F}^{-1}\left(\frac{\mathcal{F}(T_{\rm mb})}{1+\mathcal{F}(K)}\right)\;.
\end{equation}
We perform this calculation in \texttt{Python} with the unsupervised Wiener-Hunt deconvolution. This function estimates the hyperparameters automatically \citep[see ][]{den_brok_co_2022}.

\subsection{Error beam contributions for $^{13}$CO(1--0)}
We present an overview of the resulting deconvolution of the entire $^{13}$CO(1--0) SWAN 30m data-only cube in Figure~\ref{fig:errorbeam}. We compute the contribution, which is the difference between the measured and the error beam free spectrum, for each line of sight where SNR${>}3$ for $^{13}$CO(1--0). The error beam contribution is negligible in the center and along the spiral arms, being 2--3\%. The value is elevated to around 5--10\% in the interarm region. This is expected because when the telescope points to the interarm regions, part of the spiral arm will be covered by the side lobes, hence boosting the signal via the error beams.  In Figure~\ref{fig:errorbeam}, we also illustrate the effect for spectra extracted within six different apertures, from which four are in the interarm and two in the spiral arm region. The error beam contribution is presented by the brown spectrum in each panel. The percentage contribution is computed within the mask that contains the signal (illustrated in blue). With 6\% and 7\%, pointings 3 and 4 show the largest contributions in this selection of pointings.   

The error beam analysis is subject to uncertainties due to the difficulty of characterizing the variations of the beam pattern with time. However, these calculations provide a reasonable upper limit for the order of magnitude of the error beam contribution. The contribution remains ${<}10\%$ for the vast majority (i.e. 98\%) of sightlines. Within the NOEMA SWAN field-of-view, the median error beam contribution is 1.5\% per pixel.  As power within the side lobes decreases with decreasing frequency, the effect will be even less significant for the other lines, such as HCN and N$_2$H$^{+}$.

\section{Consistency tests of the archival and new SWAN 30m data sets}
\label{app:archival_vs_new30m}

For this comparison, both the SWAN and EMPIRE 30m data are processed in the same way using the EMPIRE pipeline \citep[see ][]{jimenez-donaire_empire_2019}. The raw spectra are first calibrated to antenna temperatures scale using the nearest chopper-wheel calibration scan. After this, each observed line is extracted using the CLASS package. For each individual line-of-sight where a spectrum is extracted, we subtract a zeroth-order baseline and regrid the spectrum to a $4\,\mathrm{km}\,\mathrm{s}^{-1}$ channel width and write them out as FITS tables. 
The spectra are then processed with an IDL procedure that allows us to flag and discard pathological data. 
We then fit a baseline excluding a velocity window determined from the much brighter mean CO~(1-0) emission, to avoid including channels that potentially contain signal. 
In addition, two more windows adjacent to the central one are included to fit a second-order polynomial baseline, which is then subtracted from the entire spectrum. 
The final spectra are then sorted according to their measured rms on the line-free windows, relative to the expected value from the radiometer equation, and the highest 10\% are rejected \citep[see][for a detailed description of each step in the pipeline]{jimenez-donaire_empire_2019}. Finally, all data corresponding to each spectral line are gridded into a cube. 
We then convolve each datacube to a common working resolution of 33\arcsec, using a Gaussian kernel.

\begin{figure}
    \centering
    \includegraphics[width = 1.0\textwidth]{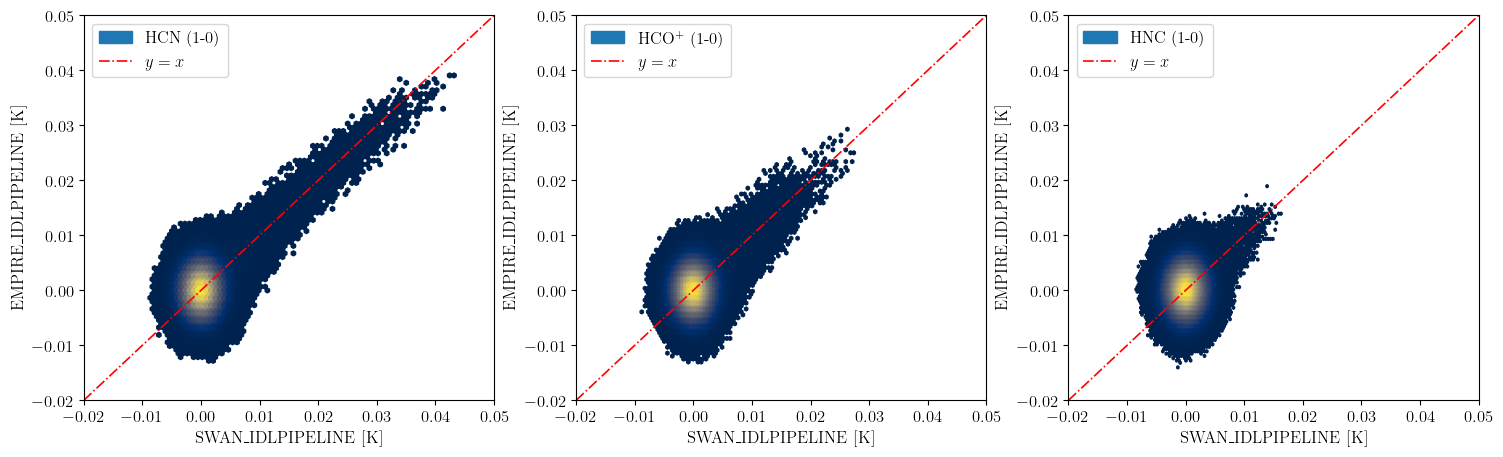}
    \caption{Pixel-by-pixel comparison of the HCN\,(1-0), \hcop\,(1-0), and HNC\,(1-0) data cubes obtained with SWAN and EMPIRE IRAM-30m observations. The red line indicates a 1-1 correlation between the EMPIRE and the SWAN datasets. Both single-dish data sets are processed with the EMPIRE pipeline for comparison and overall agree with each other.}
    \label{fig:EMPIRE_pipeline_comparison}
\end{figure}

Figure~\ref{fig:EMPIRE_pipeline_comparison} shows a comparison between the main dense gas products obtained with SWAN and EMPIRE 30m observations, processed using the EMPIRE pipeline described above. These include the main lines HCN~(1--0) (left panel), HCO$^+$(1--0) (middle panel), and HNC~(1--0) (right panel). The red line indicates a 1-1 correlation between the EMPIRE and the SWAN datasets. As can be seen from the figure, both datasets agree well overall for the three different lines. We quantify this by calculating the relative sum of all differences between both datasets. We find that the HCN\,(1--0), HCO$^+$(1--0), and HNC~(1--0) measurements agree between EMPIRE and SWAN within a 1\%, 7\% and 12\%, respectively.

\section{Comparison of moment map integration techniques}
\label{app:Momentmappipelines}

\begin{figure} 
    \centering
    \includegraphics[width = 0.5\textwidth]{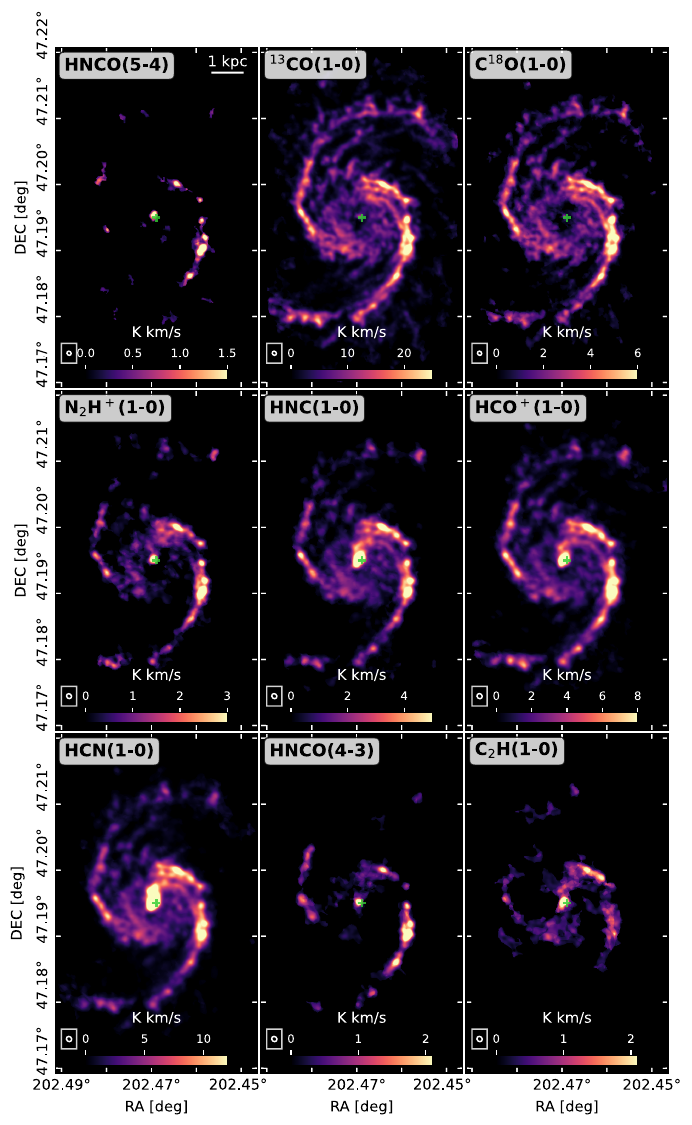}
    \caption{Integrated intensity maps (moment-0) of the SWAN data set (combined NOEMA and IRAM 30m observations) for the J=1--0 transitions of \tco, \ceto, \nnhp, \hcop, HNC, HCN, \cch{}(1--0 and HNCO(J=4--3) plus HNCO(J=5--4) at their native angular resolution ($\sim2.3-3.1\arcsec$). 
    The maps are created with the \texttt{GILDAS} ``Island-method''.  
    We show the beam size in the bottom of all panels as well as a 1 kpc scale bar in the top left panel. 
    }
    \label{fig:Gallery_GILDAS}
\end{figure}

\begin{figure}
    \centering
    \includegraphics[width = 0.5\textwidth]{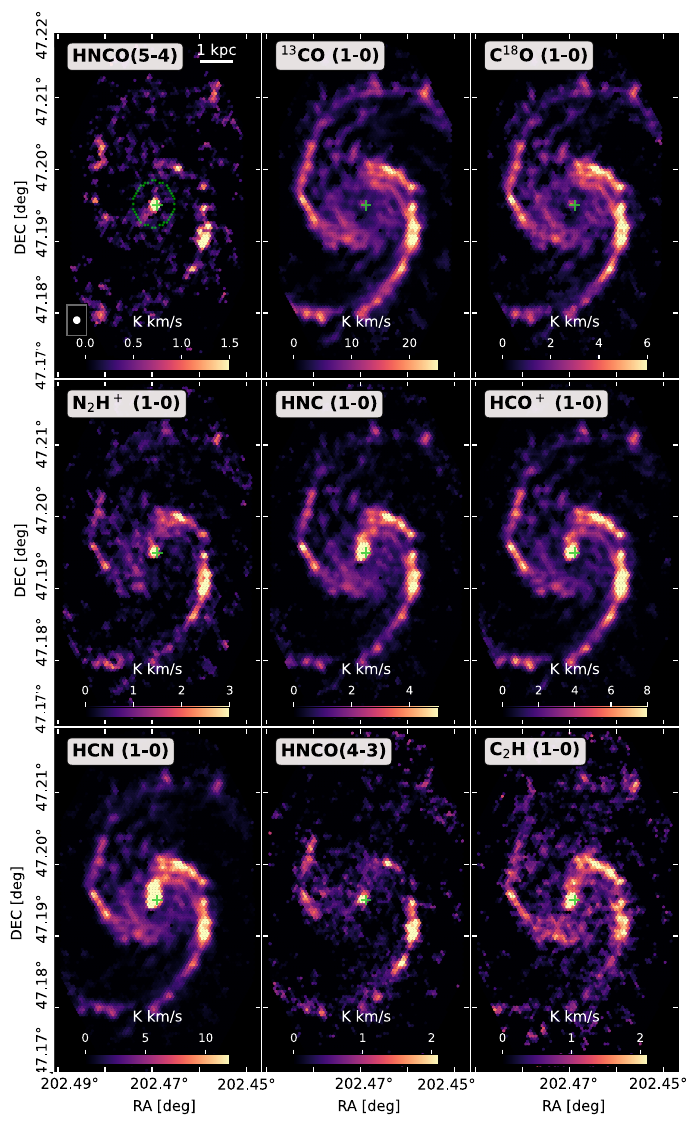}
    \caption{Integrated intensity maps (moment-0) of the SWAN data set (combined NOEMA and IRAM 30m observations) for all detected lines at a common resolution of $3\arcsec$ (125\,pc). The data is binned with hexagonal spacing with the \texttt{PyStructure} code \citep{den_brok_co_2022,neumann_almond_2023}. 
    Spectral windows for the creation of the moment maps are selected based on significant detections of \CO emission from PAWS \citep{schinnerer_pdbi_2013} and HCN(1-0) emission. 
    We show the beam size, a 1 kpc scale bar and mark the central 1\,kpc circular area (green points) in the top left panel. The intensity scale is the same as in Figure~\ref{fig:Gallery_GILDAS}. }
    \label{fig:Gallery}
\end{figure}

\begin{figure}
    \centering
    \includegraphics[width = 0.5\textwidth]{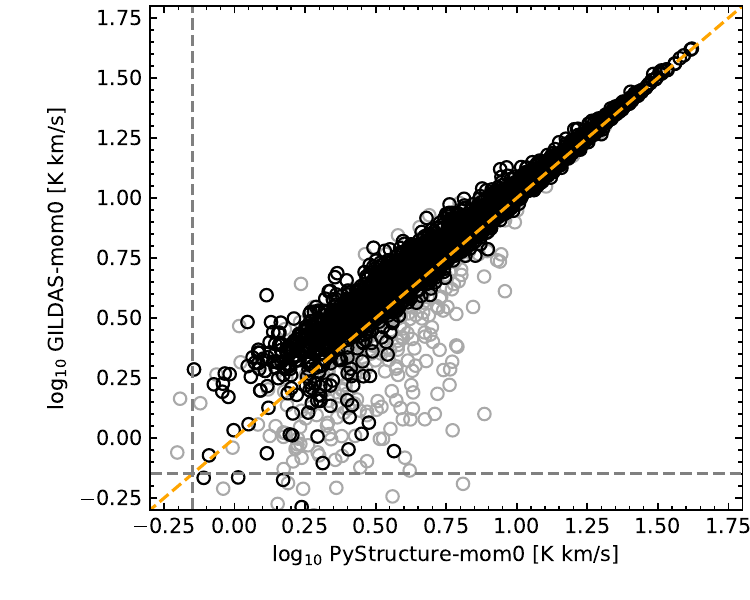}
    \caption{Pixel-by-pixel comparison of the obtained integrated line emission using two different methods for the 10\,km\,s$^{-1}$ resolution \tco data cube at native angular resolution.
    We show pixels located inside (black circles) and outside (grey circles) the hull of the mosaics (compare Figure~\ref{fig:mosaicpointings}).
    We show the 1:1 relation (orange dashed line). 
    We mark the average 5$\sigma$ noise level for both lines (grey dashed line). 
    We note that due to logarithmic spacing, data points containing noise with negative fluxes are not visible. 
    Although this applies to most data points in the interarm region near the edges of our FoV, we emphasize that this comparison is intended to assess how both methods handle regions with significant detected emission, as these areas are typically the focus of scientific analysis. 
    Regions with significantly detected emission is found mostly in the center, the molecular ring, and on the spiral arms.
    }
    \label{fig:Pystructure_vs_GILDAS}
\end{figure}

Given both the \texttt{GILDAS} and PyStructure methods are useful for different analysis, we aim to confirm a general agreement between both methods. 
To do so, we re-grid the \tco moment-0 map produced with \texttt{GILDAS} and has rectangular pixels, to match the hexagonal pixels of the moment-0 map produced with the PyStructure code. 
This is done using parts of the PyStructure code. 
Figure~\ref{fig:Pystructure_vs_GILDAS} shows the emission in each hexagonal pixel from the re-gridded \texttt{GILDAS}-moment-0 map compared to that from the moment-0 map that was inherently produced with the PyStructure code. 
The moment maps generated by the \texttt{GILDAS} method recovers about 25\% more flux than the Pystructure method (Figure~\ref{fig:Pystructure_vs_GILDAS}). 
The median flux difference between both maps is 0.9 K\,km\,s$^{-1}$ with 16$^\mathrm{th}$ and 84$^\mathrm{th}$ percentiles of 0.3 and 2.1 K\,km\,s$^{-1}$.
For pixels at high intensities, both methods agree well. 
The PyStructure method is more conservative and misses broader linewings. 
As the PyStructure table used for the scientific analysis in this paper is based on \CO and HCN as priors (see above), we do not expect to miss any significant emission.

\section{Average logarithmic line ratios in comparison}
\label{app:Offsets}

We list mean logarithmic line ratios ($b$) from Figure~\ref{fig:Lineemissioncomparison} (Section~\ref{sec:Science:Linevsline}) for all detected lines in Table~\ref{tab:offsetb}. 
$b$ is calculated between two lines (line-x and line-y) in pixels where both lines are significantly detected ($>3\sigma$). 
$b$ is defined as $b=\mathrm{mean}\left( \mathrm{log}_{10} \left(y/x\right) \right)$ with $x$ and $y$ referring to the values on the x and y-axis from line-x and line-y respectively. 
Additionally, we provide $b_\mathrm{cen}$ and $b_\mathrm{disk}$, which is the same calculation performed on pixels inside and outside the central 1\,kpc, respectively. 
To estimate how strong the central effects on the line ratios are, we calculate $\sigma_\mathrm{cen-disk}$, which is the significance of the difference of  $b_\mathrm{cen}$ and $b_\mathrm{disk}$, defined as $\sigma_\mathrm{cen-disk} = \frac{b_\mathrm{cen} - b_\mathrm{disk}}{\sqrt{ \Delta b_\mathrm{cen}^2 + \Delta b_\mathrm{disk}^2}}$.
There are only very few pixels with significant HNCO(5-4) detections, which might bias the calculation of $\sigma_\mathrm{cen-disk}$.

With this, we find the following: 
The most extreme difference ($\sigma_\mathrm{cen-disk}>50$) in mean logarithmic line ratios between center and disk can be found for HCN and \tco, as well as for HNC and \tco. 
This is followed by line ratios between HCN and \hcop, HCN and \ceto, HNC and \ceto, \hcop and \ceto, \cch and \ceto ($30 <\sigma_\mathrm{cen-disk}<50$).  
The most extreme differences are therefore seen between the \CO isotopologues and most other lines. 
This is consistent with the visual lack of \tco and \ceto emission compared to all other detected lines in the galaxy center seen in Figure~\ref{fig:Gallery}.

Most other line combinations have significantly different line ratios in the center compared to the disk, but with a lower value of $\sigma_\mathrm{cen-disk}<30$. 
For HCN, the strongest offset relation between center and disk is with \ceto emission, the weakest with isomer HNC and the faint HNCO(5-4) line. 

Line combinations that do not exhibit a clear offset relation between center and disk are \nnhp and \hcop, \ceto and \tco, and combinations including the HNCO and \cch lines. 
All line combinations with \tco show a significant offset relation between center and disk, with the only exception being the other \CO isotopologue, \ceto.

\begin{table}
\begin{small}
\caption{Average logarithmic line ratios in comparison}\label{tab:offsetb}
\centering
\begin{tabular}{ll|rrr|r}
\hline\hline
\noalign{\smallskip}
line-x & line-y & $b_\mathrm{all}$ & $b_\mathrm{cen}$ & $b_\mathrm{disk}$ & $\sigma_\mathrm{cen-disk}$\\
(1) & (2) & (3) & (4) & (5) & (6)\\
 \noalign{\smallskip}
 \hline 
\noalign{\smallskip}
\nnhp{}(1-0) & HCN(1-0) & 0.651$\pm$0.003 & 0.831$\pm$0.009 & 0.625$\pm$0.004 & 21 \\
\nnhp{}(1-0) & HNC(1-0) & 0.273$\pm$0.004 & 0.377$\pm$0.010 & 0.257$\pm$0.004 & 11 \\
\nnhp{}(1-0) & \hcop{}(1-0) & 0.488$\pm$0.003 & 0.476$\pm$0.010 & 0.489$\pm$0.004 &  1.3 \\
\nnhp{}(1-0) & HNCO(5-4) & -0.249$\pm$0.011 & -0.108$\pm$0.028 & -0.275$\pm$0.012 & 6 \\
\nnhp{}(1-0) & HNCO(4-3) & -0.141$\pm$0.007 & -0.151$\pm$0.020 & -0.140$\pm$0.007 &  0.5 \\
\nnhp{}(1-0) & \cch{}(1-0) & 0.003$\pm$0.006 & 0.093$\pm$0.015 & -0.012$\pm$0.007 & 6 \\
\nnhp{}(1-0) & \tco{}(1-0) & 0.983$\pm$0.003 & 0.771$\pm$0.009 & 1.013$\pm$0.003 & 25 \\
\nnhp{}(1-0) & \ceto{}(1-0) & 0.395$\pm$0.004 & 0.196$\pm$0.012 & 0.420$\pm$0.004 & 18 \\
 \noalign{\smallskip}
 \hline 
\noalign{\smallskip}
HCN(1-0) & HNC(1-0) & -0.397$\pm$0.002 & -0.440$\pm$0.005 & -0.391$\pm$0.002 & 9 \\
HCN(1-0) & \hcop{}(1-0) & -0.160$\pm$0.002 & -0.395$\pm$0.005 & -0.136$\pm$0.002 & 48 \\
HCN(1-0) & HNCO(5-4) & -0.872$\pm$0.010 & -1.074$\pm$0.022 & -0.831$\pm$0.011 & 8 \\
HCN(1-0) & HNCO(4-3) & -0.769$\pm$0.005 & -1.080$\pm$0.016 & -0.730$\pm$0.006 & 21 \\
HCN(1-0) & \cch{}(1-0) & -0.674$\pm$0.004 & -0.801$\pm$0.010 & -0.652$\pm$0.005 & 13 \\
HCN(1-0) & \tco{}(1-0) & 0.419$\pm$0.001 & 0.045$\pm$0.003 & 0.454$\pm$0.002 & 117 \\
HCN(1-0) & \ceto{}(1-0) & -0.201$\pm$0.002 & -0.496$\pm$0.007 & -0.173$\pm$0.002 & 46 \\
 \noalign{\smallskip}
 \hline 
\noalign{\smallskip}
HNC(1-0) & \hcop{}(1-0) & 0.219$\pm$0.002 & 0.047$\pm$0.006 & 0.242$\pm$0.002 & 29 \\
HNC(1-0) & HNCO(5-4) & -0.517$\pm$0.010 & -0.609$\pm$0.024 & -0.497$\pm$0.012 &  4 \\
HNC(1-0) & HNCO(4-3) & -0.400$\pm$0.006 & -0.608$\pm$0.016 & -0.372$\pm$0.006 & 14 \\
HNC(1-0) & \cch{}(1-0) & -0.280$\pm$0.005 & -0.335$\pm$0.011 & -0.270$\pm$0.005 & 5 \\
HNC(1-0) & \tco{}(1-0) & 0.774$\pm$0.002 & 0.469$\pm$0.005 & 0.816$\pm$0.002 & 68 \\
HNC(1-0) & \ceto{}(1-0) & 0.168$\pm$0.003 & -0.074$\pm$0.008 & 0.196$\pm$0.003 & 32 \\
 \noalign{\smallskip}
 \hline 
\noalign{\smallskip}
\hcop{}(1-0) & HNCO(5-4) & -0.740$\pm$0.010 & -0.675$\pm$0.024 & -0.754$\pm$0.011 &  3.0 \\
\hcop{}(1-0) & HNCO(4-3) & -0.620$\pm$0.005 & -0.672$\pm$0.016 & -0.614$\pm$0.006 &  3.4 \\
\hcop{}(1-0) & \cch{}(1-0) & -0.473$\pm$0.005 & -0.414$\pm$0.011 & -0.483$\pm$0.005 & 6 \\
\hcop{}(1-0) & \tco{}(1-0) & 0.567$\pm$0.002 & 0.422$\pm$0.005 & 0.582$\pm$0.002 & 30 \\
\hcop{}(1-0) & \ceto{}(1-0) & -0.043$\pm$0.002 & -0.126$\pm$0.008 & -0.036$\pm$0.002 & 11 \\
 \noalign{\smallskip}
 \hline 
\noalign{\smallskip}
HNCO(5-4) & HNCO(4-3) & 0.119$\pm$0.013 & -0.058$\pm$0.033 & 0.147$\pm$0.014 & 6 \\
HNCO(5-4) & \cch{}(1-0) & 0.126$\pm$0.014 & 0.228$\pm$0.029 & 0.104$\pm$0.015 &  3.8 \\
HNCO(5-4) & \tco{}(1-0) & 1.086$\pm$0.010 & 0.636$\pm$0.023 & 1.175$\pm$0.011 & 21 \\
HNCO(5-4) & \ceto{}(1-0) & 0.515$\pm$0.011 & 0.062$\pm$0.029 & 0.597$\pm$0.012 & 17 \\
 \noalign{\smallskip}
 \hline 
\noalign{\smallskip}
HNCO(4-3) & \cch{}(1-0) & 0.102$\pm$0.009 & 0.272$\pm$0.023 & 0.072$\pm$0.010 & 8 \\
HNCO(4-3) & \ceto{}(1-0) & 0.490$\pm$0.006 & 0.262$\pm$0.022 & 0.512$\pm$0.006 & 11 \\
HNCO(4-3) & \tco{}(1-0) & 1.058$\pm$0.005 & 0.780$\pm$0.016 & 1.092$\pm$0.005 & 18 \\
 \noalign{\smallskip}
 \hline 
\noalign{\smallskip}
\cch{}(1-0) & \tco{}(1-0) & 0.953$\pm$0.004 & 0.608$\pm$0.011 & 1.012$\pm$0.005 & 35 \\
\cch{}(1-0) & \ceto{}(1-0) & 0.384$\pm$0.005 & 0.083$\pm$0.015 & 0.422$\pm$0.005 & 21 \\
 \noalign{\smallskip}
 \hline 
\noalign{\smallskip}
\tco{}(1-0) & \ceto{}(1-0) & -0.611$\pm$0.002 & -0.626$\pm$0.007 & -0.610$\pm$0.002 & 2.3 \\
\end{tabular}
\tablefoot{Offset $b$ from Figure~\ref{fig:Lineemissioncomparison} for pixels where both lines (line-x, line-y) are significantly detected. We define $b=\mathrm{mean}\left( \mathrm{log}_{10} \left(y/x\right) \right)$ with $x$ and $y$ referring to the values on the x and y-axis from line-x and line-y respectively.  
We calculate $b$ for pixels where both line-x and line-y are significantly detected in the full FoV (3), the central 1\,kpc (4), and the remaining disk excluding the central 1\,kpc (5). We add the significance of the difference of $b$ calculated in the center compared to the disk. We define it as $\sigma_\mathrm{cen-disk} = \frac{b_\mathrm{cen} - b_\mathrm{disk}}{\sqrt{ \Delta b_\mathrm{cen}^2 + \Delta b_\mathrm{disk}^2}}$. 
Values including HNCO(5-4) should be taken with caution, as this line is only significantly detected in very few pixels in the FoV.
} 
\end{small}
\end{table}

\section{\tco line ratios}
\label{app:13coratios}

\begin{figure} 
    \centering
    \includegraphics[width = 0.48\textwidth]{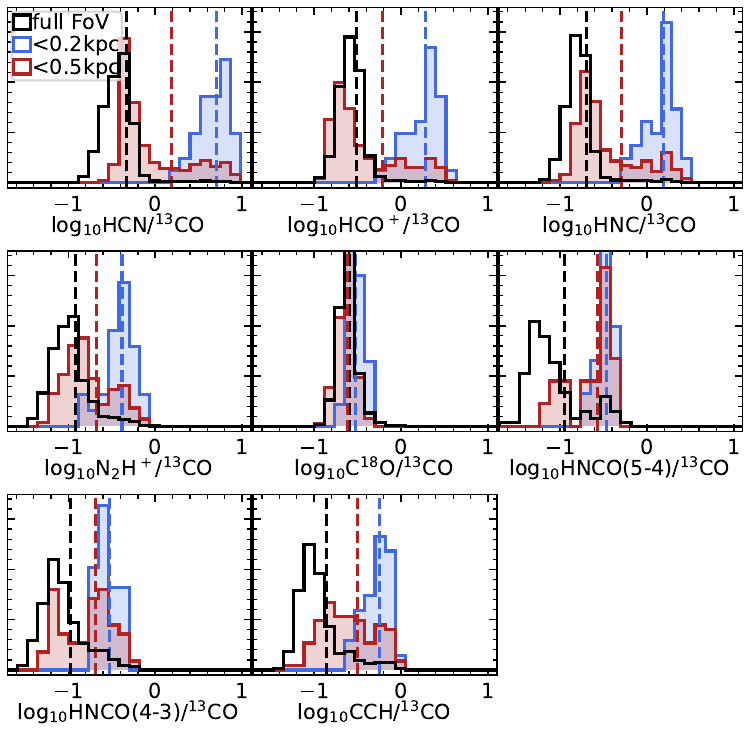}
    \caption{Same as Figure~\ref{fig:Histograms_LinevsCO} but for line ratios with \tco(1-0). }
    \label{fig:Histograms_Linevs13CO}
\end{figure}

We show histograms of line ratios with \tco emission in the full FoV, central 1\, and 0.4\,kpc (diameter) in Figure~\ref{fig:Histograms_Linevs13CO}.
In agreement with the \CO line ratios (Figure~\ref{fig:Histograms_LinevsCO}), we see the same qualitative behavior: 
Line ratios in the center are increased for all lines except the isotopologue \ceto. 
This increase is even stronger for the smaller (0.4\,kpc) aperture for HCN, HNC, \hcop, \nnhp and \cch. 
We note that the average distribution of the HNCO(5-4)/\tco line does not vary much between 1\, and 0.4\,kpc. As the HNCO(5-4) emission is only detected in few pixels in the very galaxy center, both histograms might depict the same information.

\end{appendix}
\end{document}